\documentclass[twocolumn]{aastex62}
\usepackage{graphicx}
\usepackage[normalem]{ulem}
\newcommand\redout{\bgroup\markoverwith
{\textcolor{red}{\rule[.5ex]{2pt}{0.4pt}}}\ULon}
\usepackage{xcolor}
\usepackage{threeparttable}
\usepackage{amsmath}
\newcommand{\mstar}{$M_*$}
\newcommand{\mhalo}{$M_{\rm halo}$}
\newcommand{\mstareq}{M_*}
\newcommand{\mhaloeq}{M_{\rm halo}}

\newcommand{\msun}{$M_{\odot}$}
\newcommand{\msuneq}{M_{\odot}}
\newcommand{\zsuneq}{Z_{\odot}}
\newcommand{\zsun}{$Z_{\odot}$}
\newcommand{\mbary}{$M_{\rm bary}$}
\newcommand{\mbaryeq}{M_{\rm bary}}
\newcommand{\rvir}{$r_{\rm vir}$}
\newcommand{\rvireq}{r_{\rm vir}}

\newcommand{\hone}{H~\textsc{i}}

\newcommand{\otwo}{O~\textsc{ii}}

\newcommand{\mgtwo}{Mg~\textsc{ii}}

\newcommand{\osix}{{\rm O}~\textsc{vi}}
\newcommand{\neeight}{Ne~\textsc{viii}}

\newcommand{\lya}{Ly$\alpha$}

\newcommand{\chiion}{$\chi_{\rm \neeight}$}

\newcommand{\fc}{$f_c$}

\newcommand{\rtwo}{$r_{200}$}

\newcommand{\beq}{\begin{equation}}
\newcommand{\eeq}{\end{equation}}

\newcommand{\cc}{{\rm cm}^{-3}}
\newcommand{\kms}{km s$^{-1}$}

\newcommand{\lcloud}{$L_{\rm cloud}$}
\newcommand{\tvir}{$T_{\rm vir}$}
\newcommand{\cmt}{{\rm cm}^{-2}}

\graphicspath{{./}{}}

\submitjournal{ApJL}

\shorttitle{CASBaH: Warm-hot Circumgalactic Gas Reservoirs Traced by Ne~VIII}
\shortauthors{Burchett et al.}

\begin{document}


\title{The COS Absorption Survey of Baryon Harbors (CASBaH): \\ Warm-hot Circumgalactic Gas Reservoirs Traced by Ne~VIII Absorption}

\correspondingauthor{Joseph N. Burchett}
\email{burchett@ucolick.org}

\author{Joseph N. Burchett}
\affil{University of California - Santa Cruz \\
1156 High St. \\
Santa Cruz, CA, USA 95064}

\author{Todd M. Tripp}
\affil{University of Massachusetts - Amherst \\
Amherst, MA, USA}

\author{J. Xavier Prochaska}
\affil{University of California - Santa Cruz \\
1156 High St. \\
Santa Cruz, CA, USA 95064}

\author{Jessica K. Werk}
\affil{University of Washington - Seattle \\
Seattle, WA, USA}

\author{Jason Tumlinson}
\affil{Space Telescope Science Institute \\
Baltimore, MD, USA}
\affil{Johns Hopkins University \\ 
Baltimore, MD, USA}

\author{J. Christopher Howk}
\affil{Department of Physics, University of Notre Dame,  \\
Notre Dame, IN 46556, USA}

\author{Christopher N. A. Willmer}
\affil{Steward Observatory, University of Arizona \\
Tucson, AZ, USA}

\author{Nicolas Lehner}
\affil{Department of Physics, University of Notre Dame,  \\
Notre Dame, IN 46556, USA}

\author{Joseph D. Meiring}
\affil{Texas Advanced Computer Center, University of Texas, \\ 
Austin, TX, USA}

\author{David V. Bowen}
\affil{Dept. of Astrophysical Sciences, Princeton University, \\ Princeton, NJ  }

\author{Rongmon Bordoloi}
\affil{MIT-Kavli Center for Astrophysics and Space Research \\
Cambridge, MA, USA }
\affil{Hubble Fellow}

\author{Molly S. Peeples}
\affil{Space Telescope Science Institute \\ 
Baltimore, MD, USA}
\affil{Johns Hopkins University \\ 
Baltimore, MD, USA}

\author{Edward B. Jenkins}
\affil{Dept. of Astrophysical Sciences, Princeton University, \\ Princeton, NJ }

\author{John M. O'Meara}
\affil{St. Michael's College \\
Colchester, VT, USA }

\author{Nicolas Tejos}
\affil{Instituto de F\'isica, Pontificia Universidad Cat\'olica de Valpara\'iso,\\
Casilla 4059, Valpara\'iso, Chile}

\author{Neal Katz}
\affil{University of Massachusetts - Amherst \\
Amherst, MA, USA}



\begin{abstract}

We survey the highly ionized circumgalactic media (CGM) of 29 blindly selected galaxies at $0.49 < z_{\rm gal} < 1.44$ based  on high-S/N ultraviolet spectra of $z\gtrsim1$ QSOs and the galaxy database from the COS Absorption Survey of Baryon Harbors (CASBaH).  We detect the Ne VIII doublet in nine of the galaxies,  and for gas with $N$(\neeight)~$>10^{13.3}~\cmt$ ($>10^{13.5}~\cmt$), we derive a \neeight\ covering fraction $f_{c} = 75^{+15}_{-25}\%$ ($44^{+22}_{-20}\%$) within impact parameter $\rho \leq$ 200 kpc of \mstar = $10^{9.5-11.5}$ \msun\ galaxies and $f_{c} = 70^{+16}_{-22}\%$ ($f_{c} = 42^{+20}_{-17}\%$) within $\rho \leq$ 1.5 virial radii. We estimate the mass in \neeight-traced gas to be $M_{\rm gas}(\rm\neeight) \geq 10^{9.5} M_\odot (Z/Z_{\odot})^{-1}$, or 6-20\% of the expected baryonic mass if the \neeight\ absorbers have solar metallicity. Ionizing Ne~\textsc{vii} to \neeight\ requires 207 eV, and photons with this energy are scarce in the CGM. However, for the median halo mass and redshift of our sample, the virial temperature is close to the peak temperature for the \neeight\ ion, and the \neeight-bearing gas is plausibly collisionally ionized near this temperature.  Moreover, we find that photoionized \neeight\ requires cool and low-density clouds that would be highly underpressured (by approximately two orders of magnitude) relative to the putative, ambient virialized medium, complicating scenarios where such clouds could survive.  Thus, more complex (e.g., non-equilibrium) models may be required; this first statistical sample of \neeight\ absorber/galaxy systems will provide stringent constraints for future CGM studies.

\end{abstract}



\section{Introduction}

The circumgalactic medium (CGM), the low-density plasma that envelops a galaxy, plays a crucial role in a variety of evolutionary processes that can be broadly grouped into inflows that feed star formation vs. outflows that truncate/regulate star formation. Due to its very low density, the CGM is typically characterized via absorption lines imprinted on the spectra of background sources \citep[usually quasi-stellar objects;][]{Boksenberg:1978fk}.  Initial CGM observations used the \mgtwo\ ion to probe the gas because the relatively long wavelengths of its resonance lines enabled detection from the ground at relatively low redshifts ($z\gtrsim0.2$), where associated galaxies were also observable \citep[e.g.,][]{Bergeron:1991qy}.  
After the deployment of the {\it Hubble Space Telescope} ({\it HST}) and other space-based ultraviolet (UV) observatories, subsequent studies were able to access \mgtwo\ at even lower redshifts \citep{Bowen:1995fj} as well as \hone\ and metals that reveal more highly ionized material \citep{Morris:1993kq,Lanzetta:1995rt,Tripp:1998kq,Chen:2001lr, Stocke:2006yu,Wakker:2009fr,Prochaska:2011yq,Stocke:2013mz,Tumlinson:2013cr,Nielsen:2013aa,Werk:2013qy,Liang:2014kx,Bordoloi:2014lr,Johnson:2015qv,Burchett:2016aa}.  Recent surveys have indicated that a significant fraction of an $L^*$ galaxy's baryons reside in a cool, photoionized $\sim 10^4$ K phase in the CGM \citep{Chen:2010uq, Stocke:2013mz, Werk:2014kx, Prochaska:2017aa}.  

However, the more highly ionized CGM phases have been harder to characterize.  X-ray absorption spectroscopy has effectively probed the hot ($T \gtrsim 10^6$ K) CGM of our Galaxy \citep[e.g.,][]{Yao:2008aa,Gupta:2012aa}, but large investments of X-ray telescope time have yielded only a few (controversial) detections beyond the Milky Way \citep{Yao:2012aa}. UV surveys can build statistically significant samples and have found high covering fractions (\fc ) of circumgalactic \ion{O}{6}, particularly within 300 kpc of star forming galaxies \citep{Stocke:2006yu,Wakker:2009fr,Prochaska:2011yq,Tumlinson:2011kx,Johnson:2015qv}. However, UV studies are hampered by the fact that \ion{O}{6} is one of the few ions easily observed at rest wavelength $\lambda_{\rm r} >$ 912 \AA\ with ionization potential (IP)~$\ga 100$ eV (IP$_{\rm O\, VI} = 114$ eV).  This limits the diagnostics of ionization models, particularly of the important ``warm-hot'' ($10^{5} - 10^{6}$ K) phase where gas cools rapidly \citep[e.g.,][]{Werk:2016aa,Stern:2018aa}. Indeed, discriminating between collisionally ionized and photoionized \osix\ has proved to be difficult; collisional ionization, photoionization, and more exotic processes have all shown consistency with at least some aspects of the \ion{O}{6} data \citep[e.g.,][]{Tripp:2008lr,Savage:2010aa,Werk:2016aa,Stern:2016aa,McQuinn:2018aa,Faerman:2017aa}. One aspect of \ion{O}{6} absorbers is clear, however: they trace kinematically complex and multiphase gas \citep[e.g.,][]{Savage:2005ab,Tripp:2008lr,Tripp:2011wd,Lehner:2009aa,Tumlinson:2011aa,Muzahid:2015aa,Qu:2016aa}.  

 \ion{N}{5} (IP $=77$ eV) can also probe the warm-hot CGM, but its lines are intrinsically weak and nitrogen can be highly underabundant given its nucleosynthetic origins, so \ion{N}{5} has limited usefulness in practice. Most other ions with high IPs have resonance lines primarily at $\lambda_{\rm r} < 912$ \AA, i.e.,\ below the \hone\ Lyman limit. In the Milky Way, lines at $\lambda_{\rm r} < 912$ \AA\ are obscured by the interstellar medium. However, those lines can be studied with {\it HST} in the spectra of QSOs at $z \gtrsim 0.45$ that are not blocked by \ion{H}{1} damped Ly$\alpha$ or strong Lyman limit absorbers. 

  To access transitions at $\lambda_{\rm r} <$ 912 \AA, the \emph{COS Absorption Survey of Baryon Harbors} (CASBaH) obtained high spectral resolution {\it HST} spectra of nine QSOs at 0.92 $< z_{\rm QSO} <$ 1.48 with the \emph{Cosmic Origins Spectrograph}  \citep[COS,][]{Green:2012qy} and the \emph{Space Telescope Imaging Spectrograph} \citep[STIS,][]{Woodgate:1998fj}. A key CASBaH goal is to cover the \ion{Ne}{8} $\lambda \lambda$ 770.41, 780.32~\AA\ doublet.  With IP$=207$ eV, \ion{Ne}{8} can persist in gas with $10^{5} \lesssim T \lesssim 10^{6}$ K and thereby complements \osix\ with the potential to break degeneracies in ionization models \citep{Savage:2005aa,Narayanan:2011aa,Pachat:2017aa,Rosenwasser:2018aa}. 
Coupled with a survey of galaxies in the QSO fields (Prochaska et al. 2019, submitted), CASBaH enables the investigation of warm-hot gas tracers in galactic halos and the physical origins of the highly ionized CGM \citep{Tripp:2011wd,Meiring:2013fj}.

In this work, we present the first CASBaH results on a statistical sample of \neeight\ absorbers in the CGM and derive constraints on their physical origins. Throughout, we assume H$_0=67.7$ km s$^{-1}$ Mpc$^{-1}$, $\Omega_{\rm M}=0.31$, $\Omega_{\rm\Lambda}=0.69$.

\section{Observations and Data}
A full description of the CASBaH program, including the survey design and data handling, is provided separately (Tripp et al., in preparation); here we briefly summarize some key details relevant to this paper.

\subsection{Target Selection}

One of the primary program goals is to provide an unbiased (as much as possible) survey of the Ne~\textsc{viii} doublet to constrain the warm-hot phase of galaxy halos.  The Ne~\textsc{viii} doublet wavelengths and intrinsic weakness of the lines impose specific requirements on the survey design: (1) the UV spectra must have high signal-to-noise ratios (S/N), and (2) the target QSOs must have $z \gtrsim$ 1 to probe a large enough pathlength to accumulate a statistically useful sample.  To avoid biases, it was important to not consider any information about foreground galaxies or intervening metal absorbers in the QSO spectra when we selected the targets. Thus, our selection criteria were simple: we selected the QSOs at $z_{\rm QSO}$ = 1 -- 1.5 that are brightest in the far-UV.  

However, because {\it HST} time is expensive, we imposed two additional requirements to maximize the amount of useful path for detection of Ne~\textsc{viii} from each sightline. First, we excluded QSOs with known `black' Lyman limit (LL) absorbers that completely suppress the QSO flux shortward of $(1 + z_{\rm LL}) \times 912$ \AA\ (and thereby reduce the Ne~\textsc{viii} pathlength), and second, we rejected QSOs with known broad absorption line outflows that contaminate the spectra with complicated and variable features that greatly impede (and often preclude) analysis of intervening systems.  The exclusion of sightlines with strong LLs biases the sample against absorbers with $N(\textsc{H~i}) \gg 10^{17}$ cm$^{-2}$, but this bias is unavoidable in Ne~\textsc{viii} surveys because if there is a LL that is black at 912 \AA , then it is impossible to detect the (shorter-wavelength) Ne~\textsc{viii} lines at 770.41 and 780.32 \AA\ in that absorber; the warm-hot gas in those absorbers must be probed with other techniques.  As shown in previous CGM studies, the \textsc{H~i} column density generally increases with decreasing impact parameter $\rho$ \citep{Prochaska:2017aa}, but the CGM is patchy and there is substantial scatter in $N(\textsc{H~i})$ vs. $\rho$, so this bias does not preclude detection of Ne~\textsc{viii} in the inner CGM.

With these selection requirements, the sample of UV-bright QSOs assembled by \citet{Tripp:1994aa} provided an ideal set of targets for CASBaH. The \citet{Tripp:1994aa} QSOs are among the UV-brightest known, they are in the optimal $z_{\rm QSO}$ = 1 -- 1.5 range, and importantly, this sample was assembled without any consideration of foreground absorbers or galaxies (such information was not even avaiable when that paper was constructed).  We therefore selected all of the QSOs from \citet{Tripp:1994aa} that do not exhibit black LLs.  We also excluded TON34 (which is relatively faint) from the \citet{Tripp:1994aa} sample.  This selection from \citet{Tripp:1994aa} provided seven of the target QSOs, and we similarly selected two additional sightlines (FBQS0751+2919 and LBQS1435-0134) from more recent QSO surveys \citep{White:2000aa,Hewett:1995aa}, again without giving any consideration to information about foreground systems.

\subsection{Ultraviolet QSO Spectroscopy}

For robust analysis, it is not sufficient to only survey the Ne~\textsc{viii} spectral region.  The spectra of QSOs at $z_{\rm QSO}$ = 1 -- 1.5 have a moderately high density of metal lines as well many \textsc{H~i} Lyman series lines \citep[see, e.g., Fig.1 in][]{Tripp:2013aa}.  Consequently, to robustly identify and measure the lines without confusion from imposters that mimic the Ne~\textsc{viii} doublet, it is helpful to fully record the UV spectra from the {\it HST} cutoff ($\lambda _{\rm ob}$ = 1150 \AA ) to \textsc{H~i} Ly$\alpha$ at the redshift of the QSO [i.e., from $\lambda _{\rm ob}$ = 1150 to $\lambda _{\rm ob}$ =  $(1+z_{\rm QSO})\times 1216$ \AA ]; knowledge of \textsc{H~i} Ly$\alpha$ and metal lines at longer wavelengths is crucial for robust identification of weak Ne~\textsc{viii} lines at shorter wavelengths.  

To achieve coverage of the Ne~\textsc{viii} region and the full \hone\ Ly$\alpha$ region, we observed the QSOs with the COS FUV G130M and G160M gratings, the COS NUV G185M and G225M gratings, and the STIS E230M echelle mode. We required higher S/N in the FUV to detect the weak Ne~\textsc{viii} lines, but much more modest S/N was sufficient in the NUV for detection of the stronger \textsc{H~i} (e.g., Ly$\alpha$) lines and stronger metals (e.g., \textsc{C~iii}, \textsc{C~iv}, and \textsc{O~vi}) that are shifted into the NUV at the redshifts of Ne~\textsc{viii} systems (see Tripp et al. 2019, in prep. for details). The data were reduced as described in \citet{Tripp:2001fk} and \citet{Meiring:2011fj}.  The high S/N COS FUV spectra
  fully cover the \ion{Ne}{8}\ doublet over the 
$0.48 < z_{\rm abs} < z_{\rm QSO}$ range with spectral resolution $\approx$ $18-20$ km s$^{-1}$ 
and S/N $\approx 15 - 50$ per resolution element.

\subsection{Galaxy Redshift Survey}

Our accompanying galaxy survey (Prochaska et al., submitted) yielded $\sim$6000 galaxy redshifts with typical uncertainties of 30 km s$^{-1}$.  However, many of these galaxies are at $z < 0.48$, which is below the redshift range accessible for \neeight\ in the HST/COS band.  In addition, some of the galaxies are at large impact parameters. We note that while there are 9 sight lines in the CASBaH program, we have only surveyed for galaxies at $z>0.48$ in 6 target fields due to limited telescope time and observing runs lost to poor weather.  Nevertheless, these 6 sight lines are representative of the full CASBaH sample in terms of the S/N and number of \neeight\ absorbers found therein. We classified galaxies based on star formation rate (SFR) and stellar mass (\mstar).  SFRs were derived from the galaxy spectra using the Balmer and/or [\otwo] emission lines as in \citet{Werk:2012qy} but assuming a \citet{Chabrier:2003pd} IMF.  We fitted spectral energy distributions (SEDs) to our galaxy photometry using the CIGALE software\footnote{\url{https://cigale.lam.fr/}} \citep{Noll:2009aa} to estimate \mstar, and we used the specific star formation rates sSFR$\, \equiv \,$SFR/\mstar$\, > 10^{-11}$ yr$^{-1}$ to distinguish star-forming from passive galaxies \citep[e.g.,][]{Tumlinson:2011kx}.  We calculated the virial radius of each galaxy as described by \citet{Burchett:2016aa}.

\begin{figure}
\centering
\includegraphics[width=0.5\textwidth]{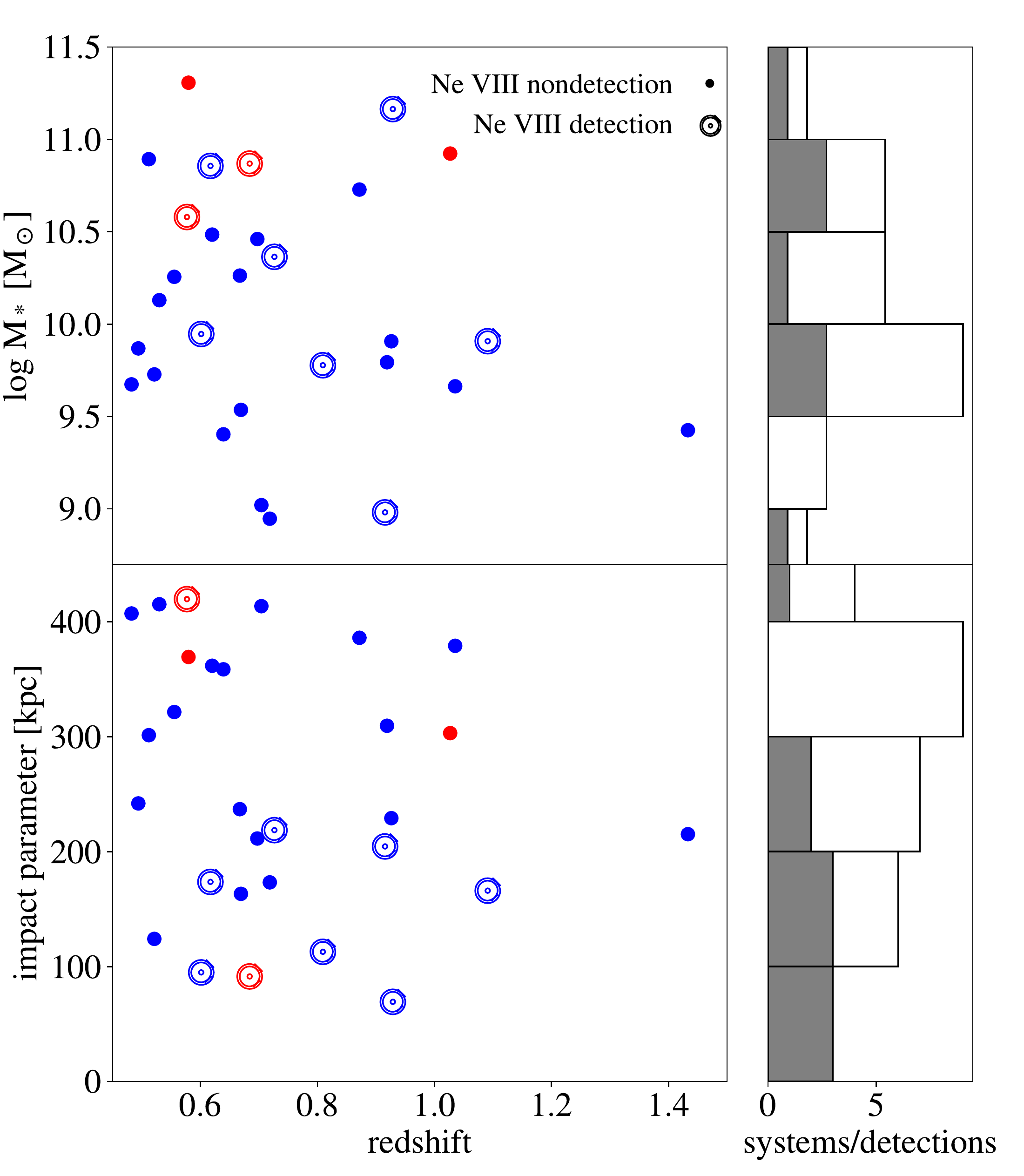}
\vspace{-0.3in}
\caption{The stellar mass (top) and impact parameter (bottom) distributions of our sample of \neeight-probed galaxies as a function of redshift; each dot represents a CGM system, and \neeight\ detections are circled.  Blue(red) symbols indicate star-forming(quiescent) galaxies.  The histograms show the number of systems in each \mstar\ and impact parameter bin with shaded regions indicating \neeight\ detections.}
\label{fig:sample}
\end{figure}

\subsection{Absorber-Galaxy Connections}

To associate our galaxy sample with absorbers (or lack thereof), we selected all galaxies from our survey with $z > 0.48$ and impact parameters $\rho<450$ kpc, or $\lesssim 1.5-4.5~\rvireq$, in the \neeight\ redshift range; the resulting sample is summarized in Table \ref{tab:sampleTable}. We note that the selection of galaxies was \textit{blind} with respect to the absorption data; i.e., no information
about the absorbers was used to assemble the sample.
In addition, the galaxies were selected for observation
without any consideration of the absorber data (Prochaska
et al., submitted). We simply targeted the brightest galaxies close to the sightlines and in the redshift range where Ne VIII can be studied. 

Independently, we assembled linelists for each sightline that included highly complete identifications and Voigt-profile fits for relevant lines.  We used the line identification procedure described by \citet{Tripp:2008lr}, which includes two passes through the data: (1) first a direct search for lines with the velocity spacing and relative strengths of the two Ne~\textsc{viii} lines, and then (2) a second pass to search for Ne~\textsc{viii} candidates that are aligned with other positively identified metals and \textsc{H~i} lines.  In the second pass, distinctive component structure in well-detected metal profiles and velocity alignment of multiple species can support the Ne~\textsc{viii} identifications (see the Appendix).  Finally, we examined the interloping lines from other redshifts that fall close to, or are blended with, the Ne~\textsc{viii} lines.  In many instances, the blended lines are constrained by features recorded elsewhere in the spectra, so the blend can be effectively modeled and removed (e.g., if the blended line is, say, an \textsc{H~i} Ly$\beta$ line, the profile of the Ly$\beta$ feature might be well-constrained by other Lyman series lines).  Because we use information on lines from other redshifts, this was an iterative procedure that ultimately relied on identification of most of the lines in the full spectra; identifications and analyses of other systems will be reported in subsequent papers. In general, we do not find compelling examples of ``isolated'' intervening Ne~\textsc{viii} absorbers; the Ne~\textsc{viii} lines are found in absorbers that exhibit other metal and \textsc{H~i} lines.  This is consistent with previous studies of intervening \textsc{O~vi} \citep[e.g.,][]{Tripp:2008lr} and Ne~\textsc{viii} systems \citep[][and references therein]{Pachat:2017aa}. In the Appendix, we provide details about interloping lines near the Ne~\textsc{viii} features presented in this paper, and we compare the apparent column-density profiles \citep{Savage:1991vn} of the two lines of the Ne~\textsc{viii} doublet as well as Ne~\textsc{viii} versus other metal profiles. Collecting the absorption components (for all species) within $\sim600$ \kms\ of one another, we cataloged absorption systems with systemic redshifts based on the central velocity of each component group. 

\begin{table*}
\caption{\neeight\ CGM sample}
\centering
\begin{rotatetable*}
\begin{threeparttable}
\renewcommand\TPTminimum{\linewidth}
\makebox[\linewidth]{
\hspace{-0.75in}
\centering
\begin{tabular}{ccccccccccc}
\hline \hline 
CGM System\tnote{a} & Sightline & RA & Dec & z$_{gal}$ & $\rho$ & log (M$_*$/M$_\odot$) & r$_{\rm vir}$ & SFR & log (N(Ne \textsc{viii})/cm$^{-2}$) & $v$\tnote{b} \\
 &  & $\mathrm{}$ & $\mathrm{}$ &  & $\mathrm{kpc}$ &  & $\mathrm{kpc}$ & $\mathrm{M_{\odot}\,yr^{-1}}$ &  & $\mathrm{km\,s^{-1}}$ \\
\hline 
J0235-0402\_121\_47 & PHL1377 & 2:35:10.1 & -4:02:29.7 & 0.5119 & 301 & $10.9 \pm  0.1$ & 242 & $ 5.7 \pm  0.1$ & $<13.63$ & -- \\
J0235-0402\_22\_56 & PHL1377 & 2:35:08.8 & -4:01:13.6 & 0.7042 & 414 & $ 9.0 \pm  0.2$ & 102 & $ 1.6 \pm  0.1$ & $<13.58$ & -- \\
J0235-0402\_221\_15 & PHL1377 & 2:35:06.8 & -4:02:16.7 & 0.8077 & 113 & $ 9.8 \pm  0.2$ & 131 & $ 0.5 \pm  0.1$ & $14.07 \pm 0.05$ & $  21 \pm   30$ \\
J0235-0402\_313\_38 & PHL1377 & 2:35:05.5 & -4:01:39.6 & 0.9189 & 309 & $ 9.8 \pm  0.2$ & 126 & $84.7 \pm  5.7$ & $<13.75$ & -- \\
J0235-0402\_360\_28 & PHL1377 & 2:35:07.4 & -4:01:37.3 & 0.9263 & 229 & $ 9.9 \pm  0.2$ & 131 & $ 3.4 \pm  0.2$ & $<13.73$ & -- \\
J0235-0402\_55\_37 & PHL1377 & 2:35:09.4 & -4:01:44.5 & 1.0269 & 303 & $10.9 \pm  0.1$ & 196 & $< 0.3$ & $<13.44$ & -- \\
J0235-0402\_14\_46 & PHL1377 & 2:35:08.1 & -4:01:21.4 & 1.0354 & 379 & $ 9.7 \pm  0.2$ & 114 & $ 8.7 \pm  0.3$ & $<13.45$ & -- \\
J0235-0402\_185\_20 & PHL1377 & 2:35:07.3 & -4:02:25.4 & 1.0894 & 166 & $ 9.9 \pm  0.2$ & 123 & $13.8 \pm  0.3$ & $14.19 \pm 0.06$ & $ -66 \pm   33$ \\
J0751+2919\_183\_39 & FBQS0751+0919 & 7:51:12.2 & 29:18:59.6 & 0.4939 & 242 & $ 9.9 \pm  0.2$ & 156 & $211.7 \pm  9.5$ & $<13.65$ & -- \\
J0751+2919\_225\_19 & FBQS0751+0919 & 7:51:11.3 & 29:19:24.5 & 0.5212 & 124 & $ 9.7 \pm  0.2$ & 146 & $ 1.9 \pm  0.3$ & $<13.30$ & -- \\
J0751+2919\_0\_64 & FBQS0751+0919 & 7:51:12.3 & 29:20:42.3 & 0.5298 & 415 & $10.1 \pm  0.2$ & 171 & $ 1.4 \pm  0.2$ & $<13.22$ & -- \\
J0751+2919\_124\_25 & FBQS0751+0919 & 7:51:13.9 & 29:19:24.2 & 0.6156 & 174 & $10.9 \pm  0.1$ & 226 & $ 1.2 \pm  0.2$ & $13.43 \pm 0.09$ & $ 196 \pm   30$ \\
J0751+2919\_29\_23 & FBQS0751+0919 & 7:51:13.1 & 29:19:58.0 & 0.6693 & 163 & $ 9.5 \pm  0.3$ & 126 & $ 1.1 \pm  0.0$ & $<13.12$ & -- \\
J0751+2919\_60\_29 & FBQS0751+0919 & 7:51:14.2 & 29:19:52.5 & 0.6973 & 211 & $10.5 \pm  0.1$ & 182 & $ 0.8 \pm  0.1$ & $<14.53$\tnote{1} & -- \\
J1151+5437\_0\_55 & PG1148+549 & 11:51:20.5 & 54:38:27.6 & 0.5796 & 369 & $11.3 \pm  0.2$ & 288 & $< 0.4$ & $<13.45$\tnote{2} & -- \\
J1151+5437\_143\_29 & PG1148+549 & 11:51:22.5 & 54:37:09.7 & 0.7249 & 219 & $10.4 \pm  0.1$ & 173 & $91.7 \pm 15.5$ & $13.68 \pm 0.07$ & $  11 \pm   30$ \\
J1208+4540\_178\_9 & PG1206+459 & 12:08:58.0 & 45:40:26.9 & 0.9273 & 69 & $11.2 \pm  0.1$ & 229 & $ 3.1 \pm  0.2$ & $14.98 \pm 0.09$ & $   5 \pm   30$ \\
J1409+2618\_188\_66 & PG1407+265 & 14:09:23.3 & 26:17:15.7 & 0.4823 & 407 & $ 9.7 \pm  0.3$ & 145 & $14.2 \pm  0.8$ & $<13.64$\tnote{3} & -- \\
J1409+2618\_245\_62 & PG1407+265 & 14:09:19.7 & 26:17:54.6 & 0.5754 & 420 & $10.6 \pm  0.2$ & 203 & $< 0.1$ & $14.01 \pm 0.05$ & $-139 \pm   31$ \\
J1409+2618\_285\_14 & PG1407+265 & 14:09:22.9 & 26:18:24.6 & 0.5998 & 95 & $ 9.9 \pm  0.2$ & 154 & $ 1.6 \pm  0.1$ & $14.18 \pm 0.05$ & $-136 \pm   33$ \\
J1409+2618\_207\_33 & PG1407+265 & 14:09:22.8 & 26:17:51.8 & 0.6674 & 237 & $10.3 \pm  0.2$ & 170 & $ 4.7 \pm  0.2$ & $<13.28$ & -- \\
J1409+2618\_72\_13 & PG1407+265 & 14:09:24.8 & 26:18:25.0 & 0.6826 & 91 & $10.9 \pm  0.1$ & 220 & $< 0.3$ & $13.49 \pm 0.10$\tnote{4} & $ -19 \pm   30$ \\
J1409+2618\_70\_49 & PG1407+265 & 14:09:27.3 & 26:18:37.4 & 0.8718 & 386 & $10.7 \pm  0.1$ & 191 & $< 0.6$ & $<12.97$ & -- \\
J1632+3737\_102\_48 & PG1630+377 & 16:32:05.1 & 37:37:39.7 & 0.5553 & 321 & $10.3 \pm  0.2$ & 178 & $< 0.4$ & $<13.55$ & -- \\
J1632+3737\_234\_52 & PG1630+377 & 16:31:57.6 & 37:37:19.6 & 0.6201 & 362 & $10.5 \pm  0.1$ & 191 & $< 0.8$ & $<13.08$ & -- \\
J1632+3737\_143\_51 & PG1630+377 & 16:32:03.7 & 37:37:09.7 & 0.6393 & 359 & $ 9.4 \pm  0.2$ & 121 & $ 3.2 \pm  0.3$ & $<13.32$ & -- \\
J1632+3737\_148\_23 & PG1630+377 & 16:32:02.2 & 37:37:30.3 & 0.7186 & 173 & $ 8.9 \pm  0.2$ & 98 & $< 0.5$ & $<13.34$ & -- \\
J1632+3737\_149\_25 & PG1630+377 & 16:32:02.2 & 37:37:28.2 & 0.9140 & 205 & $ 9.0 \pm  0.0$ & 91 & $22.8 \pm  0.2$ & $14.42 \pm 0.02$ & $  55 \pm   30$ \\
J1632+3737\_260\_25 & PG1630+377 & 16:31:59.1 & 37:37:45.7 & 1.4331 & 215 & $ 9.4 \pm  0.2$ & 89 & $ 7.8 \pm  0.3$ & $<14.03$ & -- \\
\hline 
\end{tabular}
}
\begin{tablenotes}
\item[a] Each CGM system name includes the underscore separated sightline name, position angle (N through E) of the associated galaxy, \\ and impact parameter in arcseconds.\item[b] The velocities given are calculated between the strongest \neeight\ component and associated galaxy.\item[1] The spectral regions where both Ne VIII $\lambda$ 770, 780 \AA\ would fall are severely affected by blending with interlopers.
\item[2] The would-be location of Ne VIII $\lambda$ 770 \AA\ is blocked by Galactic geocoronal Lya.
\item[3] The Ne VIII $\lambda$ 770 \AA\ for this system is not covered by the data due to its redshift being too low.
\item[4] The same \neeight\ component is associated with both of these CGM systems, as our system matching algorithm associated this component \\ with two separate absorber systems due to the clustered components at similar redshifts over a wide velocity range.
\end{tablenotes}
\label{tab:sampleTable}
\end{threeparttable}
\end{rotatetable*}
\end{table*}

Finally, we crossmatched the blindly selected galaxies with absorption systems from the independent line lists assembled for each sightline, defining CGM systems where galaxy-absorption velocity separations were $|\Delta v|\leq 500$ \kms\ to include absorbers at or beyond the escape velocity of the more massive halos in our sample.  
If no \neeight\ line was identified in these systems, we measured the 3-$\sigma$ upper limit within $\Delta v=\pm30$ km s$^{-1}$ of the galaxy redshift using the spectrum's flux uncertainty if no interloping line fell at the would-be \neeight\ $\lambda770$ or $\lambda780$  location; in the presence of interloping absorption, we assigned the apparent optical depth-estimated column density as the conservative upper limit.  

Figure \ref{fig:sample} shows the impact parameter, stellar mass, and redshift distribution of our resulting sample.  The absorption profiles of our CGM \neeight\ detections are presented in Figure \ref{fig:stackplots} along with other species within each system to highlight the coinciding component structure corroborating each detection. These additional species will be further analyzed in subsequent publications, although we include apparent column density profiles of our \neeight\ lines juxtaposed with other species in the Appendix as further corroboration.  Also in the Appendix, we conservatively assess the probability of our \neeight\ sample being contaminated by \lya\ from other redshifts to be $\sim0.5\%$

In cases where multiple galaxies had $\rho<450~$kpc and were at similar redshift ($|\Delta v| \lesssim 500$ \kms), we chose the galaxy with the smallest impact parameter as the reference galaxy for the CGM system.  In principle, one could use different criteria.  The most massive galaxy in a group near a sightline might better reflect the mass (and thus extent, virial temperature, etc.) of the larger dark matter halo being probed \citep{Bordoloi:2011uq,Moster:2013lr}, while that with the smallest impact parameter might dominate its local vicinity or sub-halo  probed by the sightline.  Of the 29 systems probed here, only 4 have 
this ambiguity under these selection methods; the remainder have the
most massive galaxy at the smallest impact parameter of the galaxies surveyed. In 3 of the 4 potentially ambiguous systems, the most massive and closest-$\rho$ galaxies have similar masses (within a factor of 2) and/or similar impact parameters ($\Delta\rho < 50$ kpc) to one another, both have $\rho > 200$ kpc and $>1.5~\rvireq$, and we do not detect \neeight\ absorption at the corresponding redshifts.  Therefore, our covering fraction calculations within 200 kpc, mass estimates, and related results are not affected by our selection criteria.  In the final ambiguous case, a \neeight\ detection at $z\sim 0.68$, the smallest-$\rho$ galaxy has $\rho\sim 140$ kpc and $\mstareq \sim 10^{10.4} \msuneq$ while the most massive has $\rho\sim 390$ kpc and $\mstareq \sim 10^{11.5} \msuneq$.  Thus, our selection of the $\rho\sim 140~$kpc galaxy as the reference does affect the $<200$ kpc covering fraction.  These instances illustrate the possibility that several of our CGM systems may reside in group environments.  While dense environments will impact the physical conditions of CGM gas \citep[e.g.,][]{Burchett:2018aa}, investigation of this issue is beyond the scope of this paper.

\section{Results}
\label{sec:results}

\subsection{Ne VIII column density profile}
\label{sec:ne8profile}

Figure \ref{fig:ne8profiles} plots the \ion{Ne}{8} column densities versus impact parameter for star-forming and passive galaxies with red and blue squares, respectively.  Also shown are the covering fractions ($f_c$) of \neeight\ absorbers with a detection threshold of log~$(N($\neeight$)/ \cmt ) = 13.3$ in bins of impact parameter.  Systems with upper limits above this threshold value are omitted from the calculation and are marked with faint squares. Figure \ref{fig:covFrac} shows $f_c$ for different threshold choices:  log~$(N($\neeight$)/ \cmt ) = 13.3, 13.5, {\rm and}~14.0$.  The completeness and depth of our survey varies from field to field, and we qualify that our $f_c$ calculations are most relevant for $\gtrsim 0.1~L^*$ galaxies. Lower-mass galaxies with smaller impact parameters may be present in certain systems, and several \neeight\ absorbers have been identified in the CASBaH data but do not have galaxy associations in our database.  Our experiment is the following: given a sample of blindly selected galaxies, what are the properties of the CGM traced by \neeight\ in those galaxies as a function of impact parameter? This is analogous to the approach used in the COS-Halos survey \citep{Tumlinson:2013cr}.

\begin{figure*}
\centering
\includegraphics[width=1\linewidth]{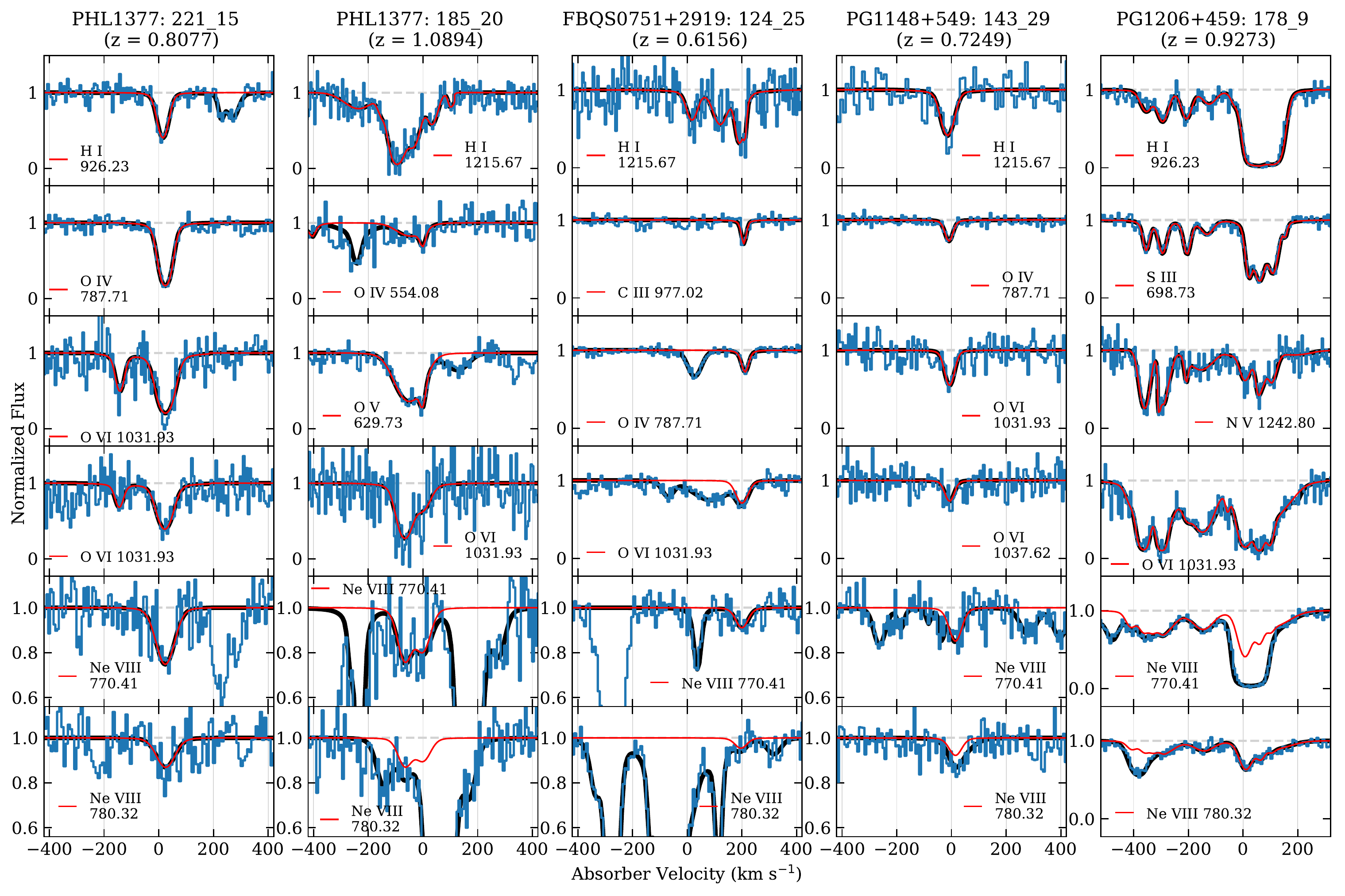}
\caption{The detections of \neeight\ absorbers in our sample along with their fitted Voigt profiles (red curves).  The velocity scale in each panel is expressed relative to the systemic redshift of the associated galaxy.  In addition to \neeight\, we show the profiles of selected other species also detected in these absorbers that corroborate the \neeight\ identification; the corresponding Voigt profiles fitted to these components are also plotted in red.  Any absorption blended with and in the vicinity of \neeight\ was fitted simultaneously with \neeight, and the full profiles are shown in black.}
\label{fig:stackplots}
\end{figure*}

\renewcommand{\thefigure}{\arabic{figure} (continued)}
\addtocounter{figure}{-1}
\begin{figure*}
\centering
\includegraphics[width=1\linewidth]{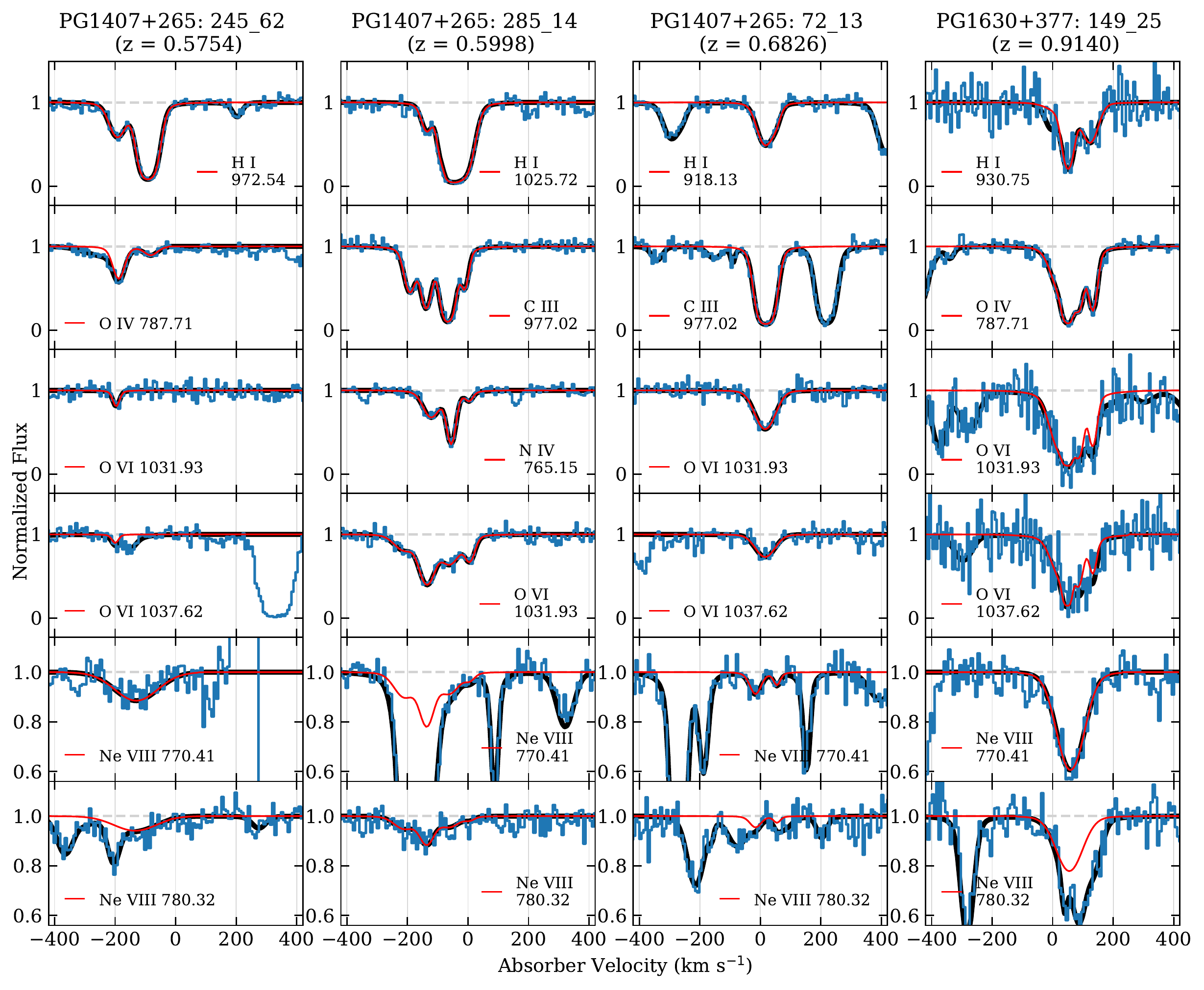}
\caption{}
\end{figure*}
\renewcommand{\thefigure}{\arabic{figure}}

\begin{figure*}
\centering
\includegraphics[width=1\linewidth]{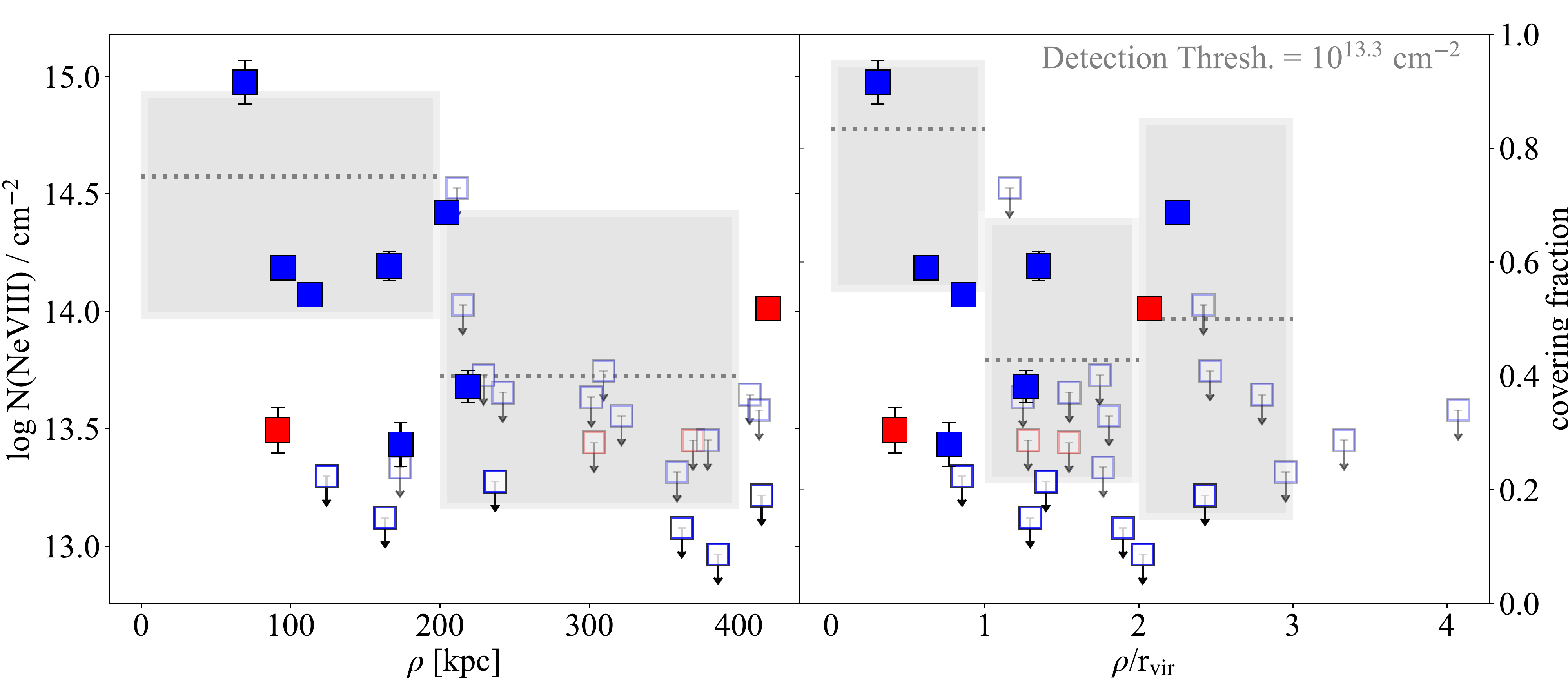}
\caption{Total \neeight\ column density per CGM system as a function of galaxy-QSO impact parameter ($\rho$; {\it left panel}) and impact parameter normalized by galaxy virial radius 
({\it right panel}). Filled squares indicate \neeight\ detections and open symbols denote upper limits, while red and blue colors represent passive and star forming galaxies, respectively.  Using the vertical scale on the right axis, \ion{Ne}{8} covering fractions ($f_c$) in bins of 150 kpc ({\it left}) or $r_{vir}$ ({\it right}), assuming a detection threshold of $N$(\neeight)$ = 10^{13.3} \cmt$, are shown by dotted lines with shaded 68\% confidence intervals about $f_c$.  Fainter markers indicate nondetections with upper limits above this threshold.}
\vspace{0.3in}
\label{fig:ne8profiles}
\end{figure*}

Within 200~kpc, we find $f_c$(\ion{Ne}{8})  =  $75^{+15}_{-25}\%$ and $44^{+22}_{-20}\%$ for log~(N(\neeight)/$\cmt$) $> 13.3$ and $>13.5$, respectively.  For the lower N(\neeight) absorbers, we measure $f_c >50\%$ within each 100 kpc bin at $\rho<~200~$kpc, albeit with large uncertainties.  Furthermore, $f_c$ declines beyond $\rho \sim 200$ kpc, although most significantly for the strongest N(\neeight) $>10^{14.0}~\cmt$ systems.  The 200---300 kpc range includes several weak upper limits due to interloping absorption that obscures \neeight\ associated with these galaxies. 

\begin{figure}
\centering
\includegraphics[width=0.5\textwidth]{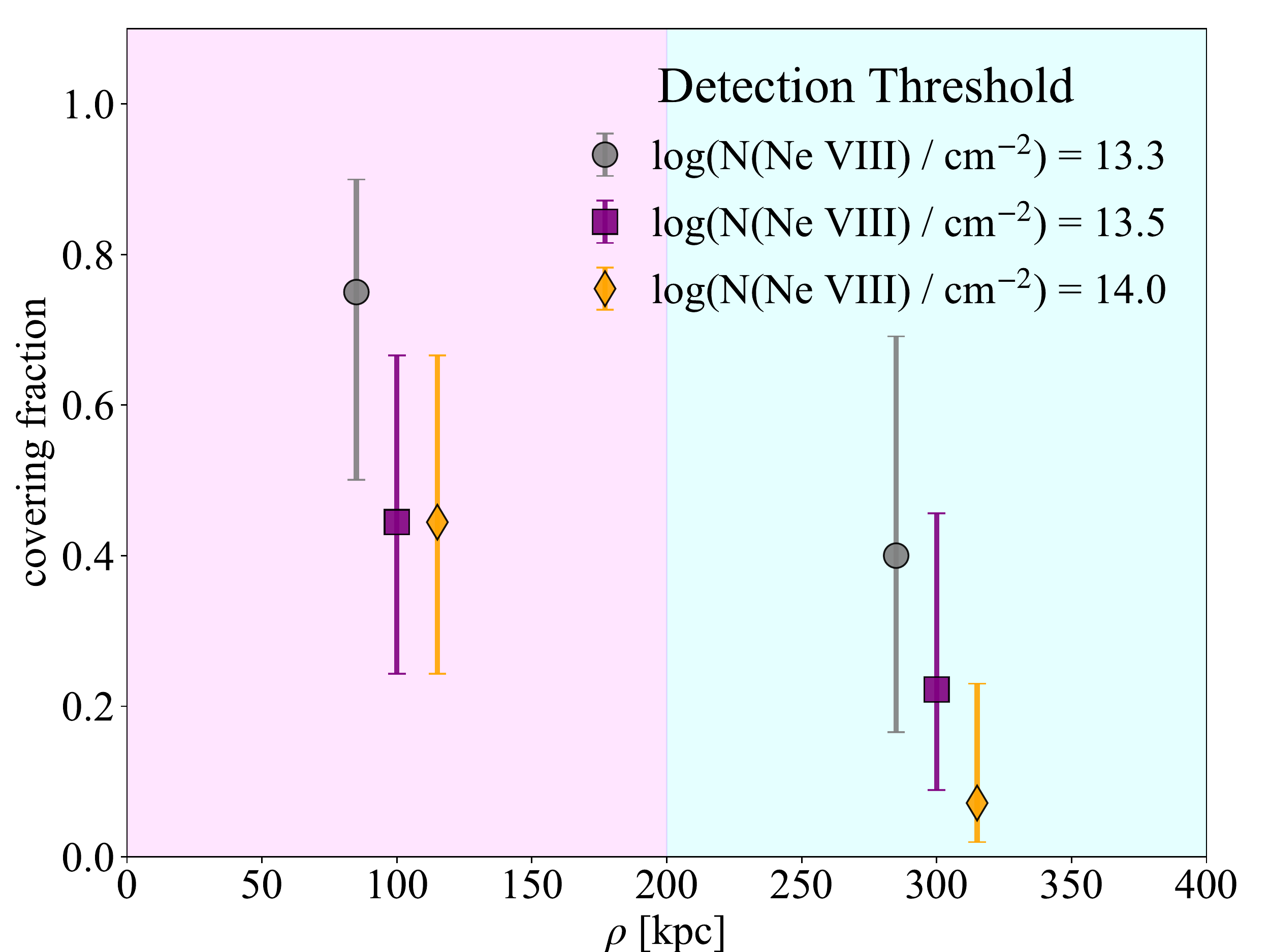}
\vspace{-0.2in}
\caption{The covering fraction of \neeight\ in two bins of impact parameter for three choices of N(\neeight) detection threshold, as indicated in the legend.  Within 200 kpc, we find $f_c$(\ion{Ne}{8})  =  $f_{c} = 75^{+15}_{-25}\%$ for a threshold of log~(N(\neeight)/$\cmt$) $= 13.3$ and $f_{c} =44^{+22}_{-20}\%$ for log~(N(\neeight)/$\cmt$) $= 13.5$ or  $14.0$.  The highest N(\neeight) absorbers show a significant decline at $\rho>200$ kpc, with $f_c = 0.07^{+0.16}_{-0.05}$. }
\label{fig:covFrac}
\end{figure}

The \neeight\ in our CGM sample spans 2 orders of magnitude in column density. Among our detections, 5/10 systems have $N$(\neeight)$>10^{14.0}~\cmt$. These $N$(\neeight) values are comparable to those measured for other metal species in the CGM, such as \osix\ \citep[e.g.,][]{Wakker:2009fr,Prochaska:2011yq, Tumlinson:2011kx, Johnson:2015xy}. 
If the \neeight\ traces gas at $T=10^{5-6}~$K, these high column densities suggest large quantities of circumgalactic gas in a phase hotter than that traced by low and intermediate ions \citep{Stocke:2013mz,Werk:2014kx,Prochaska:2017aa} at a temperature comparable to the virial temperature.    
\subsection{Mass of the \neeight\ phase}

We now derive a conservative constraint on the implied mass in the \neeight\ phase by combining reasonable assumptions about the ionization fraction of \neeight\ and the gas metallicity with our measured covering fraction, column densities, and impact parameter statistics:

\begin{equation}
\label{eq:ne8mass}
\begin{split}
\centering
M_{\rm gas}(\rm \neeight) =\  & 10^{9.5} M_\odot \left( \frac{N(\rm \neeight)}{10^{13.3}~\cmt}\right)  \left( \frac{R_{\rm CGM}}{200~\rm kpc}\right)^2 \\
& \times \left( \frac{f_c}{0.75}\right) \left( \frac{\chi_{\rm \neeight}}{0.23}\right)^{-1} \left( \frac{Z}{Z_\odot}\right)^{-1} ,
\end{split}
\end{equation}

\noindent where $R_{\rm CGM}$ is an assumed radius of the \neeight-traced gas, $f_c$ is the covering fraction of \neeight\ at the specified column density within that radius, $\chi_{\rm \neeight}$ is the ionization fraction \neeight/Ne, 
and $Z_\odot$ is the solar metallicity.  

We note that $10^{9.5} M_\odot$ is a conservative lower limit for several reasons.  First, the $\chi_{\rm \neeight}$  value 
we adopt is the peak ion fraction for \neeight\ assuming collisional ionization models \citep{Gnat:2007fk}; we include photoionization models in the next section.  Our assumption of solar metallicity is a limiting case.  \citet{Prochaska:2017aa} find that only 5/27 of the COS-Halos galaxies with well-constrained CGM metallicities have $Z\geq Z_{\odot}$ in the cool, photoionized phase of their halo gas.  
Because the median redshift of our sample of 29 CGM systems is $z=0.68$, 
it is unlikely to be more metal rich than that of the $z\approx0.2$ COS-Halos galaxies if metal enrichment increases with cosmic time, 
given that our sample spans a similar range of stellar masses. As the $M_{\rm gas}(\rm \neeight)\propto~Z^{-1}$, subsolar metallicities will increase this mass estimate substantially, by $\gtrsim3\times$($10\times$) for 0.3(0.1) $Z_\odot$.
Finally, we used a fiducial value $N$(\ion{Ne}{8}) = $10^{13.3}~\cmt$ to obtain $10^{9.5} M_\odot$; this is the lowest \ion{Ne}{8} column that we measure, and in several of the detections $N$(\ion{Ne}{8}) is substantially higher; using the median detected $N$(\ion{Ne}{8}) (10$^{14.07}$ cm$^{-2}$) and corresponding $f_c$ ($\sim$40\%) would yield $3\times$ higher mass.

Assuming the stellar/halo mass relation of \cite{Moster:2013lr}, at the median redshift and stellar mass of our $\rho<200$~kpc sample, the cosmic baryon fraction and halo mass are 0.11 and $10^{11.7}M_{\odot}$, respectively.  Thus, the CGM gas mass implied by our \neeight\ measurements 
comprises at least $\sim6-20$\% of 
the total baryonic mass (\mbary) expected for these galaxies.    

\subsection{The physical conditions of Ne VIII-traced gas}
\label{sec:ne8physical}
\begin{figure*}
\centering
\includegraphics[width=1\linewidth]{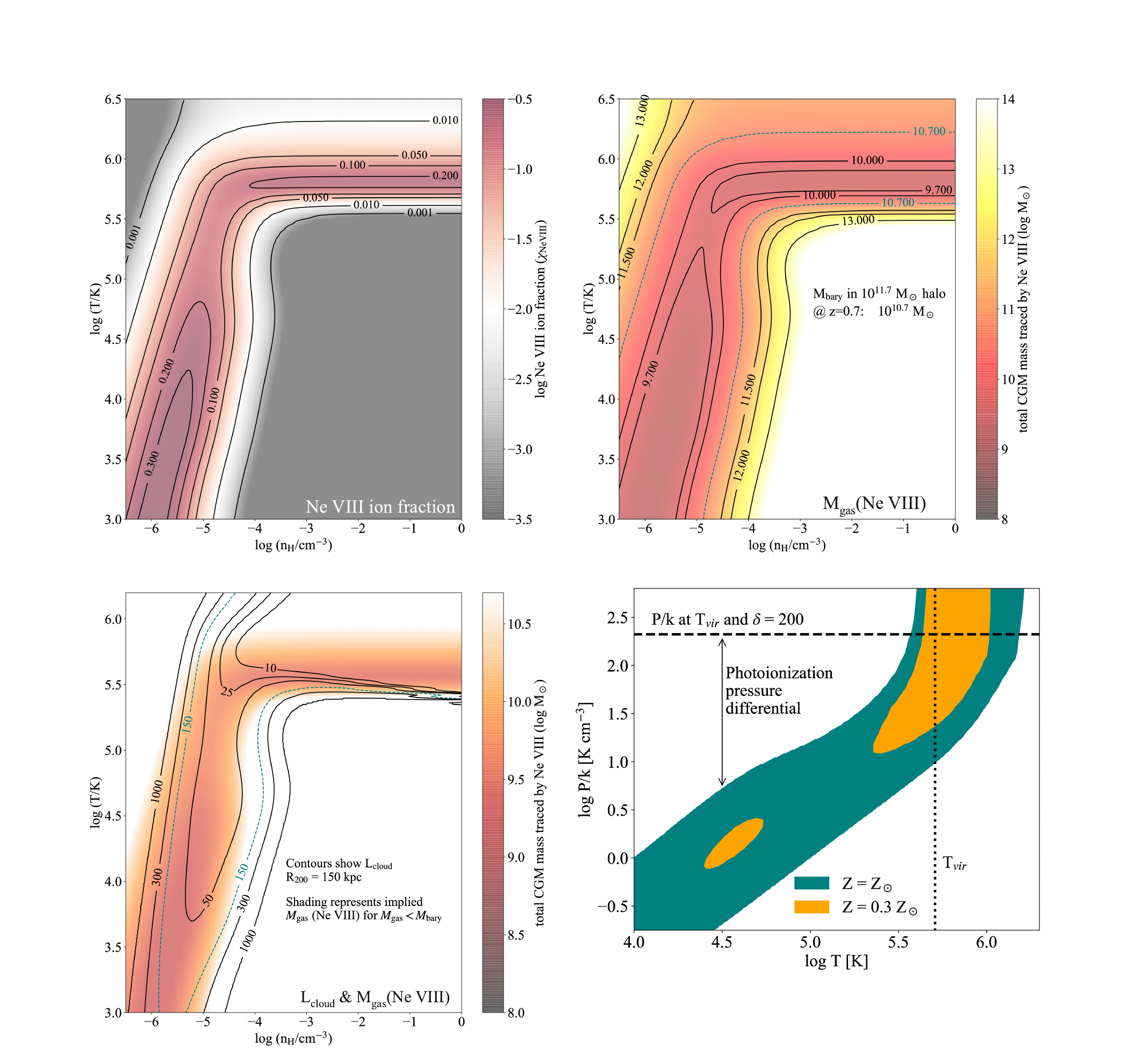}
\caption{\emph{Top-left:}  The \neeight\ ionization fraction (\chiion)  across a range of temperatures and densities, as calculated from Cloudy photoionization models assuming $Z = \zsuneq$.  \chiion\ values are both color coded and indicated by the contours.  Peaks in \chiion\ are found in both collisional ionization and photoionization dominated regimes at $\log T/{\rm K}$ $\sim5.8$ and $<4.3$, respectively, and $\log n_{\rm H}/\cc$ $>-4$ and $<-5.2$, respectively.  
\emph{Top-right:} Implied total mass traced by \neeight\ across phase space using Eqn. \ref{eq:ne8mass} and \chiion\ values from the top left panel.  The baryonic mass, assuming the cosmic fraction, of our sample's median mass ($\log \mhaloeq/\msuneq = 11.7$, $\log \mbaryeq/\msuneq = 10.7$) is marked with a teal, dashed contour; we posit that the \neeight-traced mass will not exceed \mbary\ for these CGM.  \emph{Bottom-left:}  The implied characteristic scale in kpc (or `cloud size') of the \neeight-traced material in contours superimposed over the implied mass, which is color coded and fades to white for implied masses $\geq \mbaryeq$.  The dashed, teal contour indicates \rvir\ of a $10^{12}~\msuneq$ halo. We adopt these fiducial limits to bracket the range of density and temperature of the \neeight-traced gas and note that both the implied masses and cloud sizes 
will increase as $(Z/Z_\odot)^{-1}$ for subsolar metallicities. \emph{Bottom-right:} Thermal gas pressures given the range of density and temperature constrained by assuming limits on the characteristic scale $L_{\rm cloud} \leq~\rvireq$ and total mass 
in the \neeight-traced phase  $M_{\rm gas}(\rm \neeight) \leq 10^{10.7}~\msuneq$ according to the ionization models shown in top left panel.  The teal and orange regions represent assumed metallicities of \zsun\ and 0.3 \zsun, respectively.  The horizontal dashed line is the pressure expected at the \tvir\ of a $10^{11.7}~\msuneq$ halo with an overdensity $\delta = 200$ times the mean matter density at $z=0.7$. \tvir\ is marked with a vertical dotted line.  If the \neeight-traced material were predominately photoionized, enriched to solar metallicity, and residing 
within a CGM that is filled with virialized gas, it would be severely 
under-pressurized (by nearly two orders of magnitude).  
If the metallicity were subsolar, as was found to be typical at $z \sim 0.2$ for $L^*$ galaxies by \citet[][0.3 \zsun]{Prochaska:2017aa}, the deviation from pressure equilibrium would be exacerbated. \label{fig:ne8phasespace}  }
\end{figure*}

We now explore the possible physical conditions of the 
\neeight-traced material by employing collisional- and photoionization models across phase space to estimate total masses and characteristic lengths
for the absorbing medium.  For all calculations, we use a high-resolution grid\footnote{\url{http://trident-project.org/data/ion_table/}} of Cloudy \citep{Ferland:2013aa} ionization models provided by the Trident team \citep{Hummels:2017aa} assuming $Z = \zsuneq$ 
and the extragalactic UV background of \citet{Haardt:2012fj} at $z=0.7$; the ion fractions from this grid are shown in the upper left panel of Figure \ref{fig:ne8phasespace}. Varying metallicity has a minor impact on the equilibrium ionization balance of individual ions that we employ here.

The densities inferred from the ionization modeling imply a characteristic scale (or `cloud size'), expressed as 

\beq
\label{eq:cloudsize}
L_{\rm cloud}=\frac{N_{\rm ion}}{n_{\rm ion}},
\eeq

\noindent where $n_{\rm ion}$ refers to the volume density of a particular ionic species,
which follows from ionization models as follows:

\beq
{n_{\rm \neeight}}=n_{H}\chi_{\rm\neeight}(Z/Z_\odot)({\rm Ne/H})_\odot 
\eeq
Hence, both the mass and cloud length depend inversely on \chiion. 

The top left panel of Figure \ref{fig:ne8phasespace} shows $\chi_{\rm \neeight}$ across a wide range of temperatures and densities.  Two peaks in the ion fraction occur, corresponding to the collisional ionization- and photoionization-dominated regimes at T $\approx 10^{5.8}$ K and T $\lesssim 10^{4.5}$ K, respectively.  Using the Trident grid of \chiion\ in Equation~\ref{eq:ne8mass}, we present resulting mass estimates for the full range of phase space in the top right panel of Figure \ref{fig:ne8phasespace}.  The contours mark particular values, which can far exceed our previous estimate, even under small deviations from the temperatures and densities where \chiion\ peaks.  

In the following analysis, we adopt the expected baryonic mass (\mbary$=10^{10.7}~\msuneq$; dashed contour) of a halo with the median mass from our sample ($z=0.67$; \mstar $=10^{9.9}~\msuneq$; \mhalo $=10^{11.7}~\msuneq$) as the maximum mass for the \neeight-traced medium.  That is, we do not expect the mass traced by \neeight\ absorption to exceed \mbary. 
In the context of our simple modeling and given this constraint, the photoionized regime necessarily requires much lower densities, which translates to much larger characteristic scales
(Figure~\ref{fig:ne8phasespace}, lower-left panel). 
The absorber scales are substantially smaller if the gas is 
collisionally ionized.  
Although the implied length scale of high-ionization material can exceed the plausible size of a galaxy halo if dominated by photoionization \citep[e.g.,][]{Savage:2005aa,Stern:2016aa,Werk:2016aa,Hussain:2017aa}, 50-kpc cloud sizes can be obtained under photoionization at densities $n_{\rm H} \sim 10^{-4.5}~\cc$ when T $\sim 10^{4}$ K.  However, the implied masses at this temperature exceed \mbary\ for modest increases in density, and the cloud sizes approach and quickly exceed 100 kpc for significantly lower densities, disfavoring a photoionization origin. \cite{Tepper-Garcia:2013ab}, employing cosmological simulations as well as an analytical model assuming equilibrium conditions, study the physical conditions of gas giving rise to \neeight\ absorption, finding consistent temperatures and densities to those we derive for absorbers of similar strength.

We now impose our fiducial mass and size constraints and consider the implications for CGM gas within the range of temperature and density meeting these criteria.  Figure \ref{fig:ne8phasespace} (bottom-right panel) shows the thermal pressure, expressed as $P/k = 2.3 n_{H} T$, over this `allowed' phase space, as a function of temperature.  
The teal region corresponds to solar metallicity gas ($Z =$ \zsun), while the orange corresponds to $Z=0.3$ \zsun, the median value inferred for cold, dense CGM gas in the COS-Halos galaxies \citep{Prochaska:2017aa}.  Note that the metallicity enters both Equations \ref{eq:ne8mass} and \ref{eq:cloudsize} for mass and \lcloud, respectively.   Therefore, subsolar metallicities increase both of our inferred quantities, resulting in narrower ranges of implied pressures at all temperatures.  For reference, we have also marked the virial temperature (\tvir) and expected pressure of the ambient halo medium assuming \tvir\ of our median halo and the average overdensity within the virial radius (we adopt \rvir = \rtwo).  For photoionized gas, the \neeight\ absorbing medium is highly underpressured \emph{by nearly two orders of magnitude}.

\section{Concluding Remarks}
Our CASBaH dataset exhibits a high covering fraction of \neeight\ to the sensitivity limits of our QSO spectra ($f_c = 83^{+12}_{-29}\%$ within \rvir, $f_c = 75^{+15}_{-25}\%$ within 200 kpc for log(N(Ne VIII)/cm$^{-2}$) = 13.3; $f_c = 50^{+25}_{-25}$\% within $r_{\rm vir}$, $f_c = 44^{+22}_{-20}$\% within 200 kpc for  log(N(Ne VIII)/cm$^{-2}$) = 13.5).   Assuming a stellar mass function from \citet{Moustakas:2013aa}, and integrating over galaxy stellar masses commensurate with our sample out to their virial radii, we estimate an absorber incidence per unit redshift of $d\mathcal{N}/dz \sim 5$, in apparent tension with that derived by \citet{Frank:2018aa}, who used a novel `agnostic-stacking' technique to estimate the incidence of \neeight\ absorbers. However, our $d\mathcal{N}/dz$ estimate is roughly consistent with that of the CASBaH \neeight\ blind absorber survey statistics (Tripp et al., in preparation) where no galaxy information was considered. In the agnostic-stacking method, all lines are assumed to be the Ne~\textsc{viii} 770.409 \AA\ transition and are stacked (regardless of known/unknown identification of a given line), and then the agnostic stack is checked for the signal of the corresponding Ne~\textsc{viii} 780.324 \AA\ line.  Of course, the Ne~\textsc{viii} 780.32 \AA\ will be greatly diluted in the agnostic stack because most of the lines that are stacked on are, in fact, {\it not} Ne~\textsc{viii} 770.41. Consequently, the method depends critically on simulating how the dilution from unrelated lines would impact the 780.32 \AA\ signal, which is difficult to do given the tremendous number and variety of lines in the CASBaH spectra. Further exploration of this tension is warranted, as it may be mitigated by, e.g., more accurately constraining the non-\neeight\ absorption-line distributions for producing the simulated agnositic stacks, which are crucial for interpreting the stacked real data.

The mass implied by our data for the \neeight-traced CGM conservatively accounts for 6\% of the galaxies' baryonic mass.  This estimate assumes the peak of \chiion\ under collisional ionization, or $T \sim 10^{5.8}$ K.  Interestingly, this is similar to \tvir\ for the median galaxy mass in our sample.  Thus, the virialized medium itself could produce significant \neeight\ absorption and likely contributes to the signature we detect. Independently, the absorber components' Doppler $b$ parameters are consistent with this temperature range; we defer the analyses of individual components to a subsequent work wherein we simultaneously employ \osix\ component fitting results.  Given the uncertainties in metallicity we have discussed above, the mass traced by \neeight\ could easily account for $20\%$ of \mbary\ or more, much of this also likely traced by \osix\ as well given the frequent \neeight-\osix\ alignment in velocity (Tripp et al., in preparation).  

We further argue that practical constraints on absorber size and mass render a photoionized origin unlikely for the \neeight\ systems, as such a medium would be highly underpressured relative to the ambient virialized halo.  In addition to the metallicity uncertainties, we note that our assumed UV background \citep[][HM12]{Haardt:2012fj} also factors heavily into the calculations. Several authors have argued in favor of higher intensity at energies above 1 Ryd than that of HM12 \citep[e.g.,][]{Khaire:2019aa}.  This effectively increases the corresponding $n_{\rm H}$ for a given \chiion\ and thus the pressure for a given temperature in Figure \ref{fig:ne8phasespace}.  
Even a factor of two increase in the HM12 background intensity (and thus $P$) 
would not fully alleviate the pressure differential under photoionization. However, our calculations only include equilibrium ionization models; e.g., the models of \citet{Oppenheimer:2013aa} include repeated on-off periods of AGN activity, elevating \neeight\ ion fractions for a given density by photoionization for several Myr after the AGN has turned off.  If our CGM \neeight\ systems were subject to repeated AGN irradiation on timescales of $<60$~Myr, \chiion\ may be increased by an order of magnitude for $n_H>10^{-4}$~cm$^{-3}$, alleviating the pressure nonequilibria present under photoionization equilibrium.

\citet{Stern:2018aa} propose a model where at $z\sim0.2$, \osix\ may trace low-$T$, low-$n$ photoionized gas beyond the accretion shock.  This argument applied to our \neeight\ data would avoid tensions arising from our simplified pressure equilibrium arguments, but the fact remains that \tvir\ is approximately equal to the peak temperature for \chiion, and an appreciable column of enriched virialized gas should produce a detectable \neeight\ signal.  In principle, the \neeight\ profile linewidths could lend further insight to the gas temperature, but the COS spectral resolution is inadequate for this purpose.  In future work, we will leverage the additional myriad CGM diagnostics afforded by CASBaH, including joint \neeight/\osix\ analyses as well as the low and intermediate ions.  Moreover, ongoing surveys such as QSAGE, MUSEQubes, and CUBS, will provide additional insights on galaxy evolution at $z\sim1$.

\section{Acknowledgements}
This research received financial support from NASA programs HST-GO-11741, HST-GO-13846, and HST-AR-14299 from the Space Telescope Science Institute, which is operated by the Association of Universities for Research in Astronomy, Inc., under NASA Contract NAS5-26555, as well as NASA ATP 80NSSC18K1016.  The conclusions of this work are based on data collected from observatories at the summit of Mauna Kea. The authors wish to recognize and acknowledge the very significant cultural role and reverence that the summit of Mauna Kea has always had within the indigenous Hawaiian community. We are most fortunate to have the opportunity to conduct observations from this mountain.  This study is also partly based on data acquired using the Large Binocular Telescope (LBT). The LBT is an international collaboration among institutions in the US, Italy, and Germany. LBT Corporation partners are the University of Arizona, on behalf of the Arizona university system; Istituto Nazionale do Astrofisica, Italy; LBT Beteiligungsgesellschaft, Germany, representing the Max Planck Society, the Astrophysical Institute of Postdam, and Heidelberg University; Ohio State University, and the Research Corporation, on behalf of the University of Notre Dame, the University of Minnesota, and the University of Virginia.

\appendix

\begin{figure*}
\centering
\includegraphics[width=0.49\linewidth]{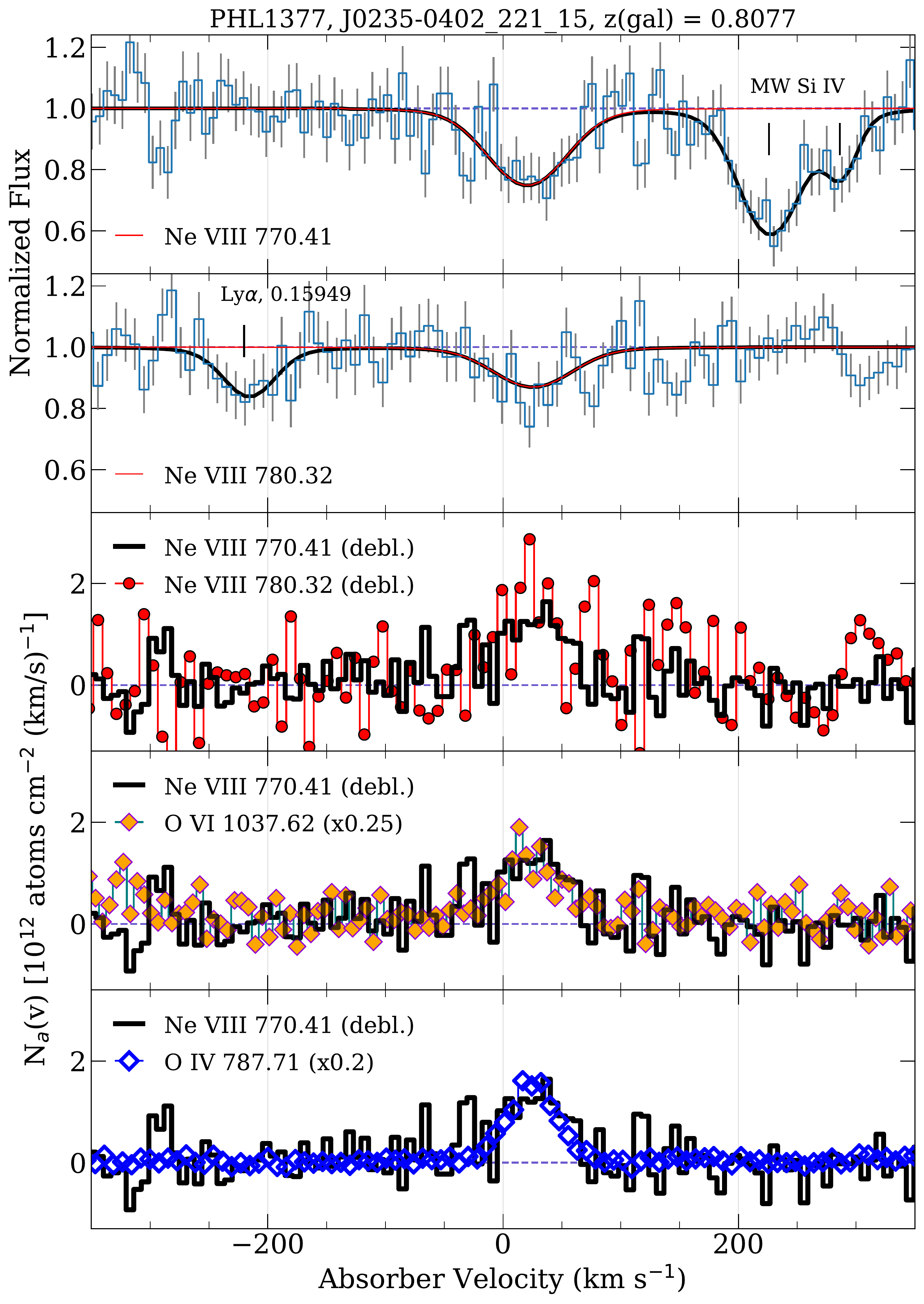}
\includegraphics[width=0.49\linewidth]{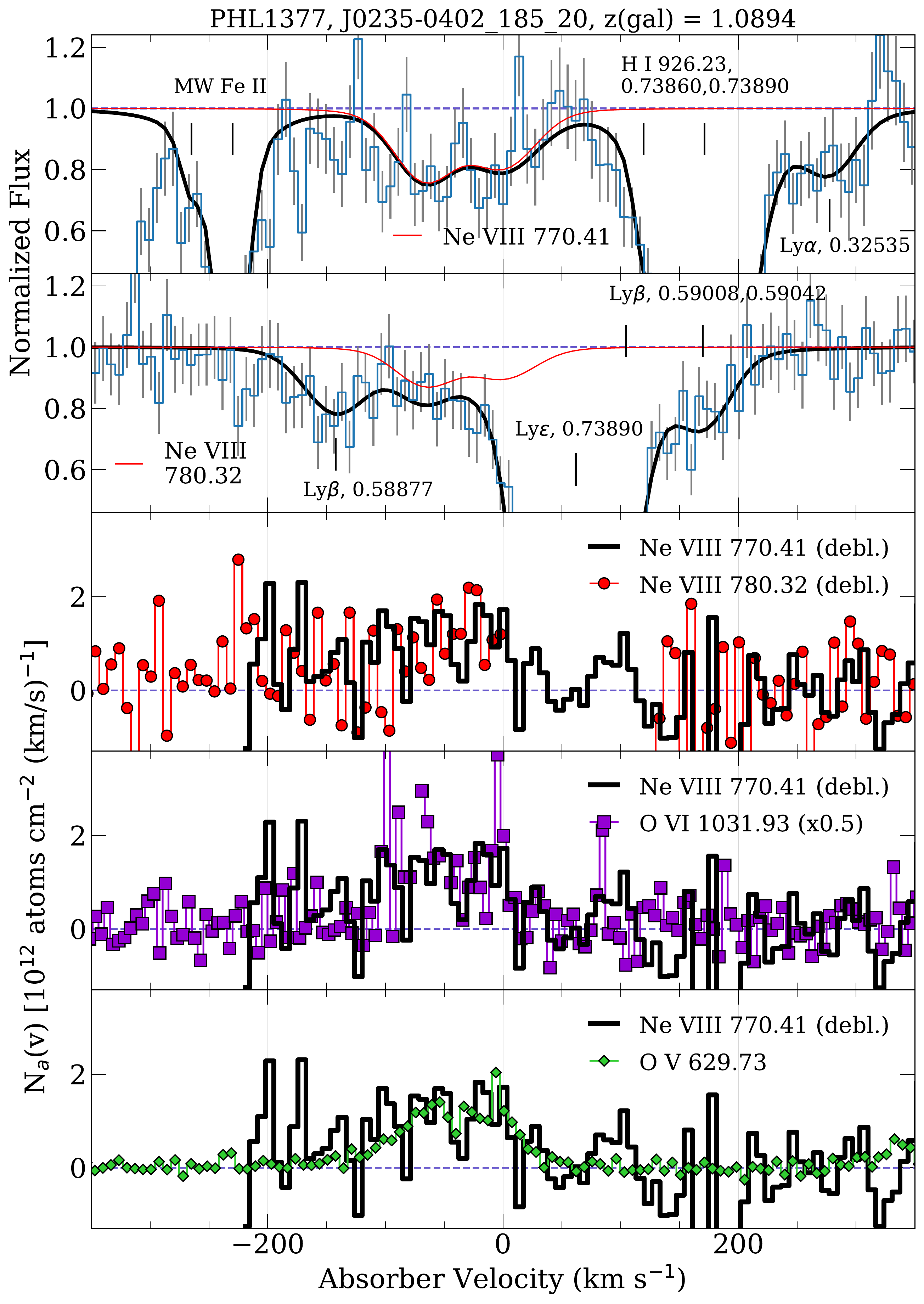}
\caption{ \neeight\ absorption line profiles (top two panels) and their apparent column density profiles juxtaposed with those from other species (remaining panels).  The corresponding sightline, CGM system name, and redshift are indicated at the top of each subfigure.  In the top two panels, the red curves mark the Voigt profile fitted \neeight\ absorption and the black curve represents the composite Voigt profile fitted to all identified lines in this spectral region (including blends).  The third panel down shows the apparent column density per pixel of each line in the \neeight\ doublet. In the remaining panels, black histograms represent the \neeight\ apparent column density per pixel, while other symbols and colors denote transitions from other species (indicated in the legend of each panel) identified at the same redshift.  Profiles marked ``debl." indicate that the optical depth from other species has been removed (deblended). \label{fig:ne8evidence}}
\end{figure*}

\setcounter{figure}{5}    
\begin{figure*}
\centering
\includegraphics[width=0.49\linewidth]{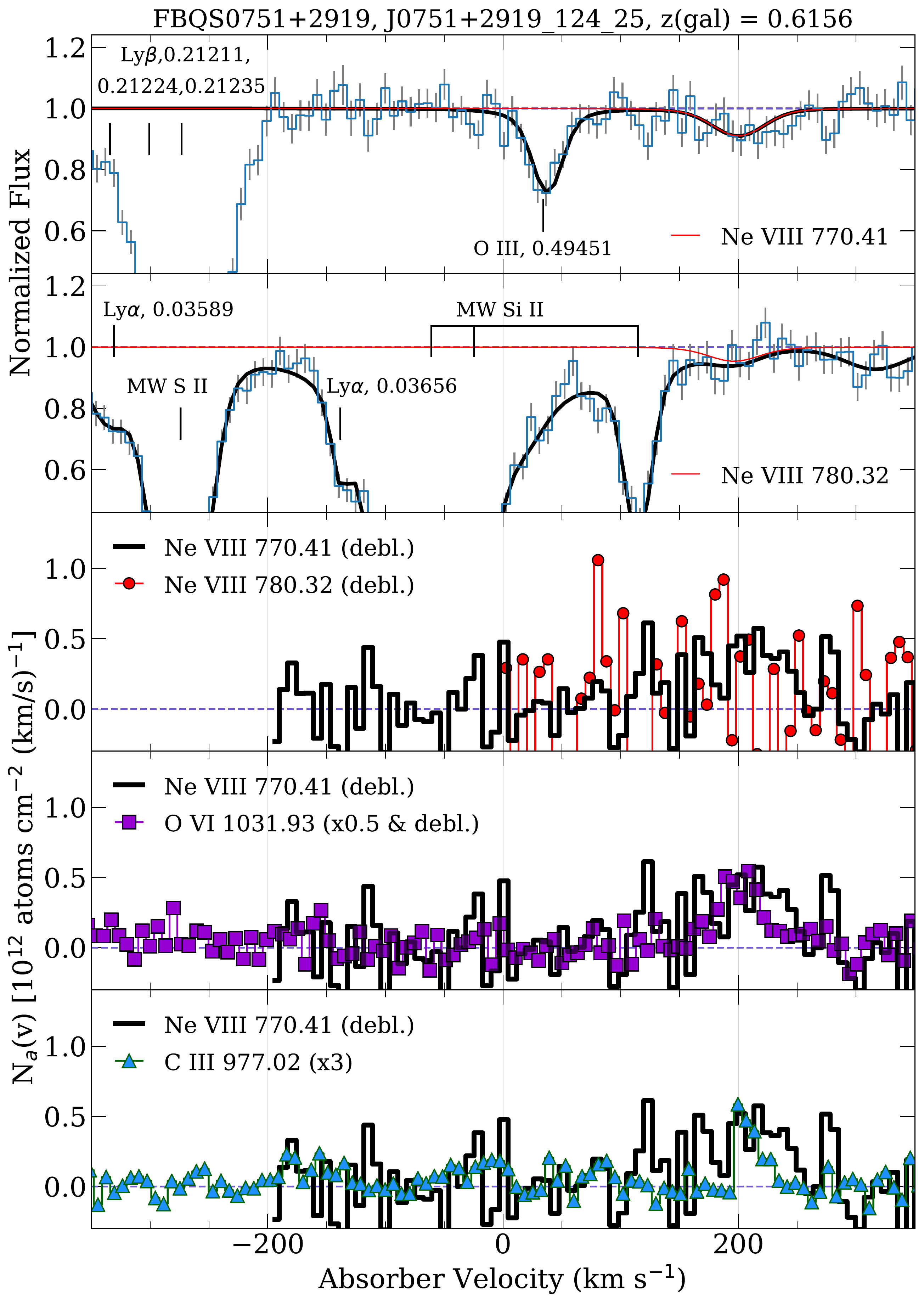}
\includegraphics[width=0.49\linewidth]{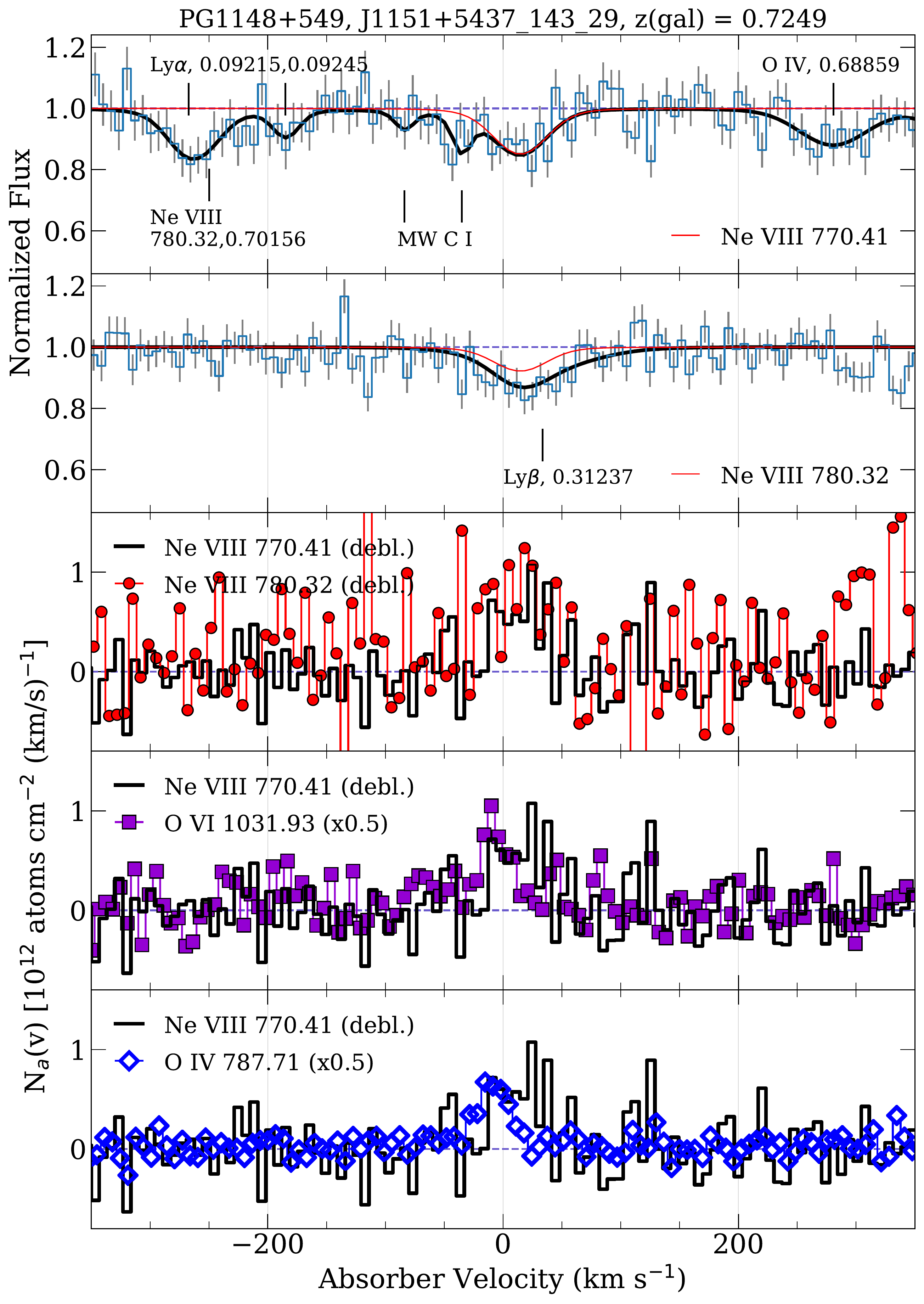}
\caption{continued}
\end{figure*}

\setcounter{figure}{5}
\begin{figure*}
\centering
\includegraphics[width=0.49\linewidth]{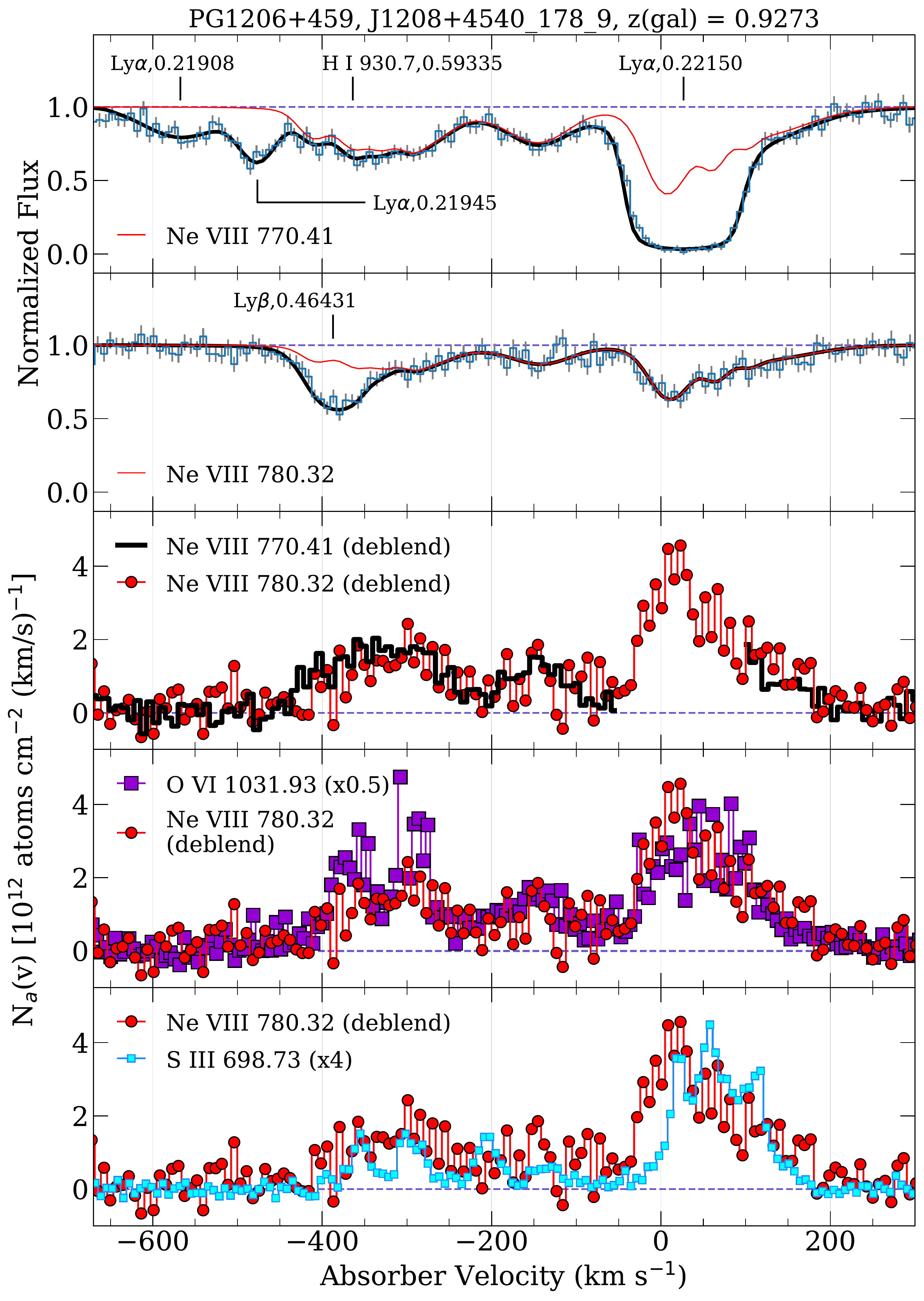}
\includegraphics[width=0.49\linewidth]{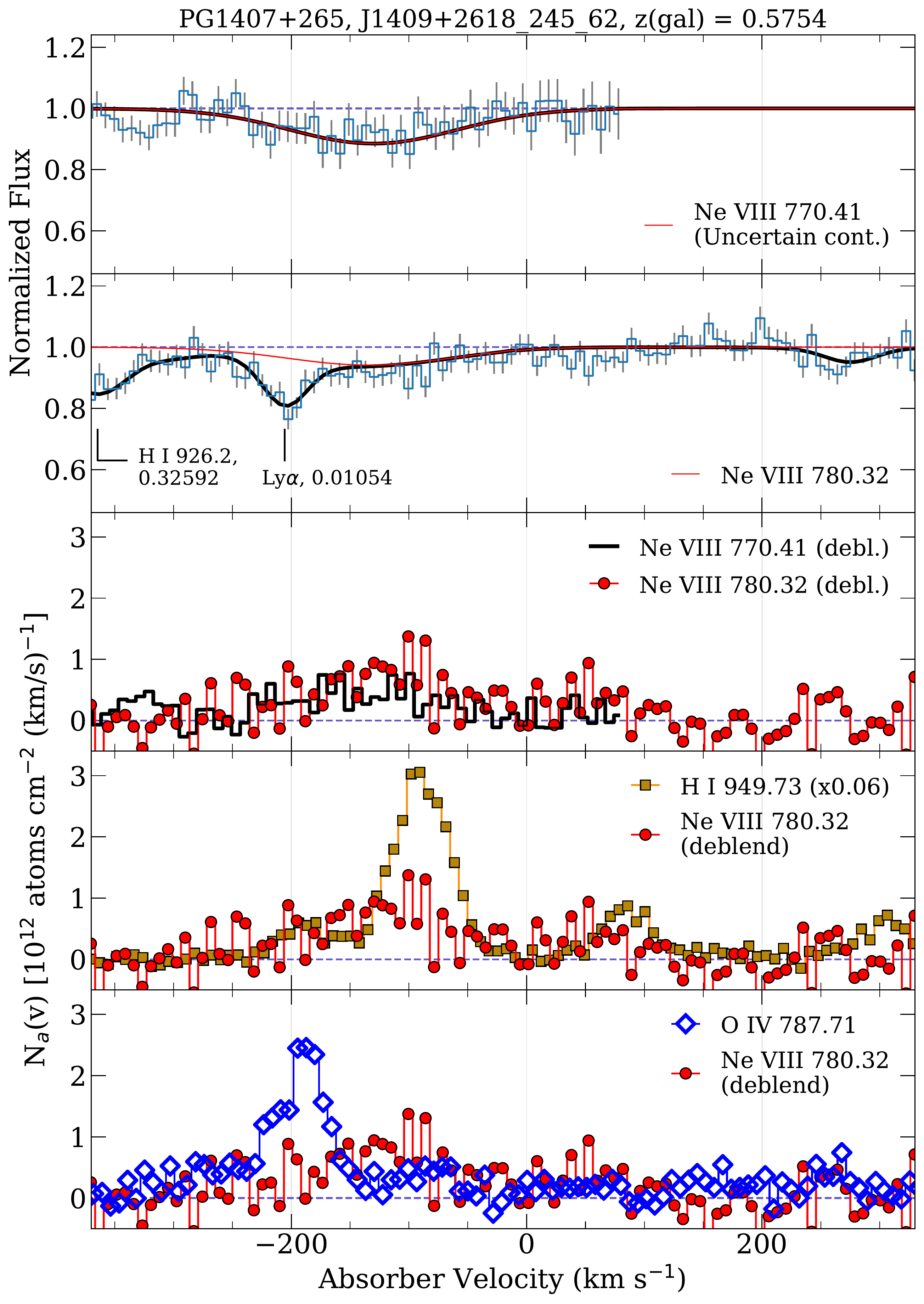}
\caption{continued}
\end{figure*}
\setcounter{figure}{5}
\begin{figure*}
\centering
\includegraphics[width=0.49\linewidth]{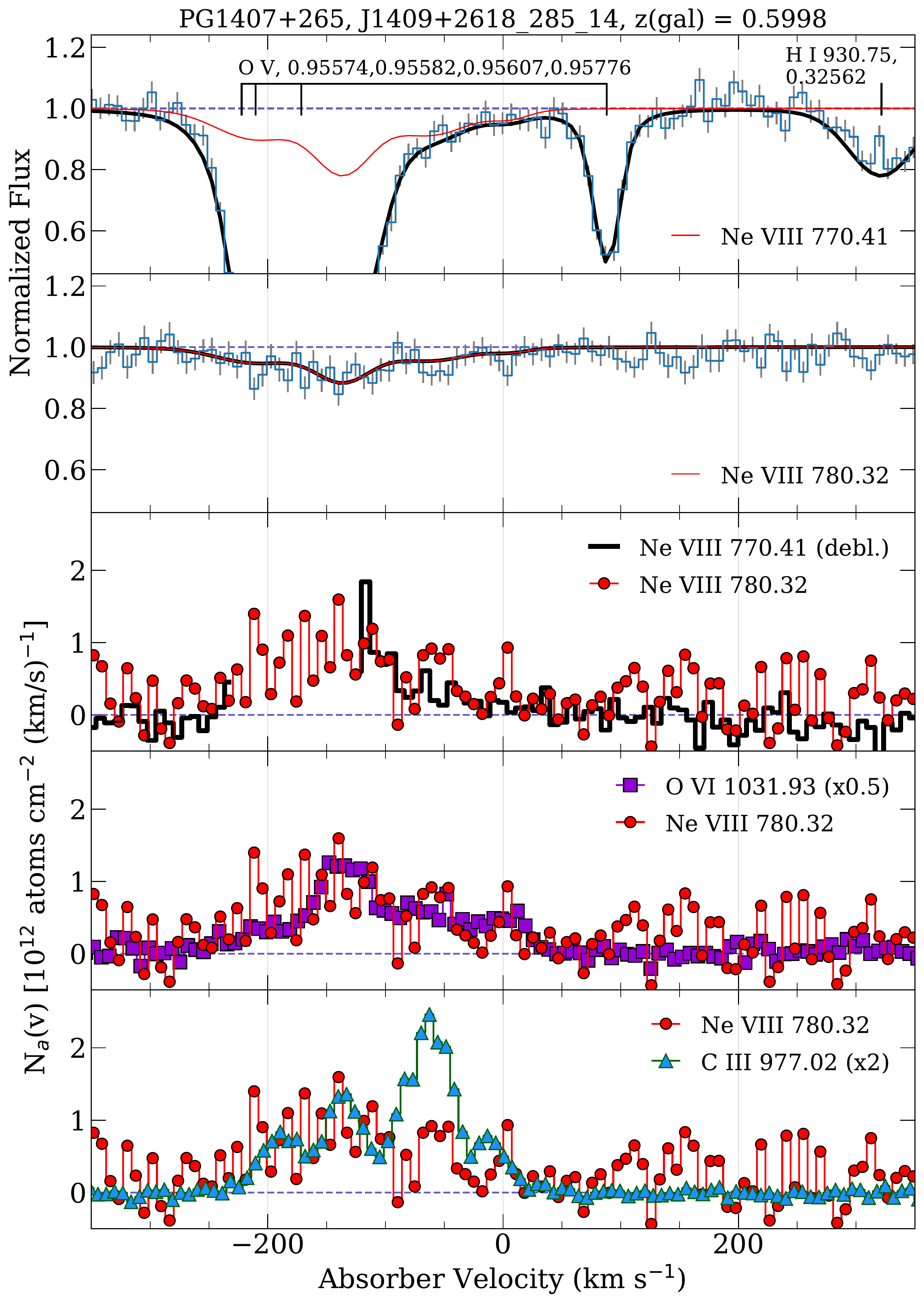}
\includegraphics[width=0.49\linewidth]{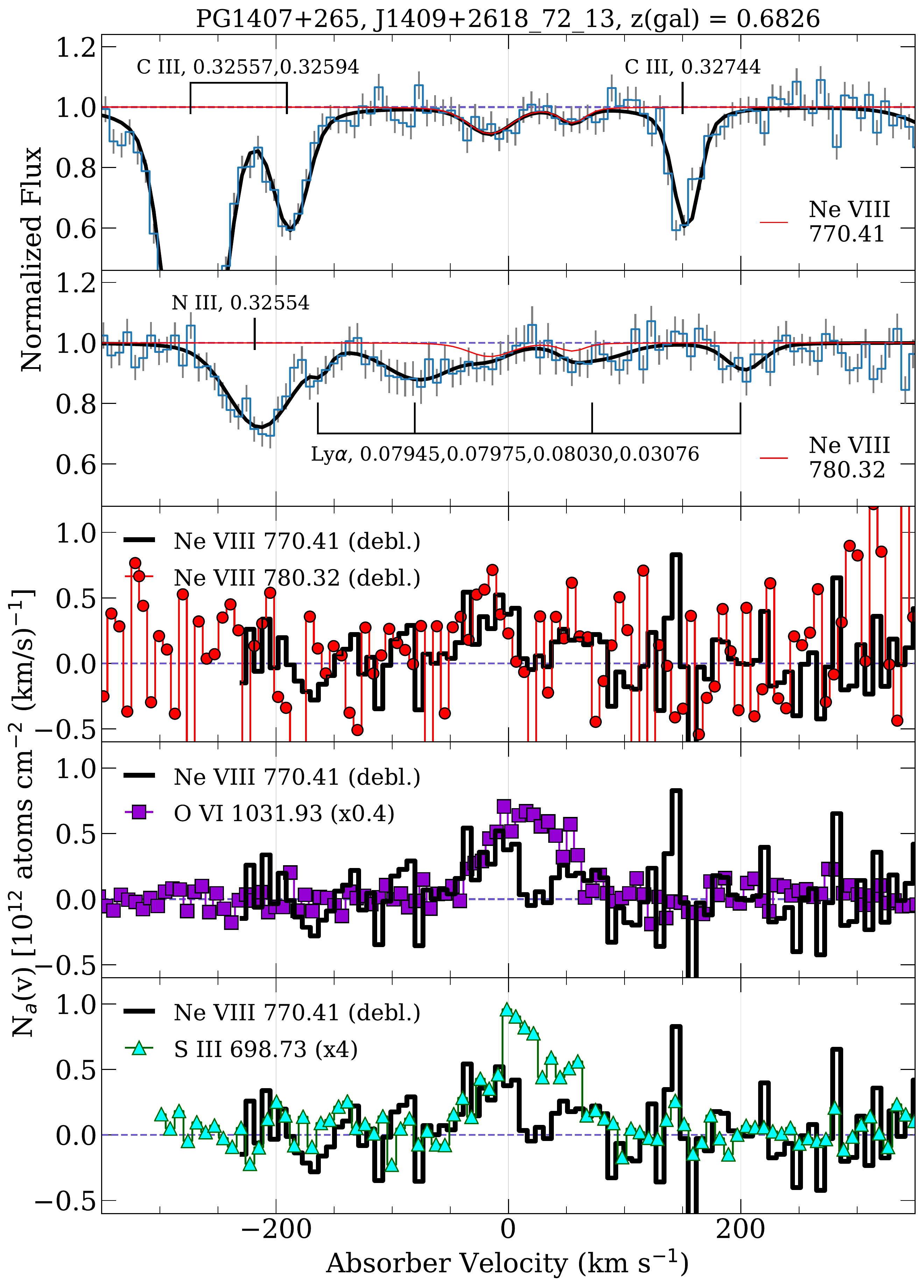}
\caption{continued}
\end{figure*}
\setcounter{figure}{5}
\begin{figure*}
\centering
\includegraphics[width=0.49\linewidth]{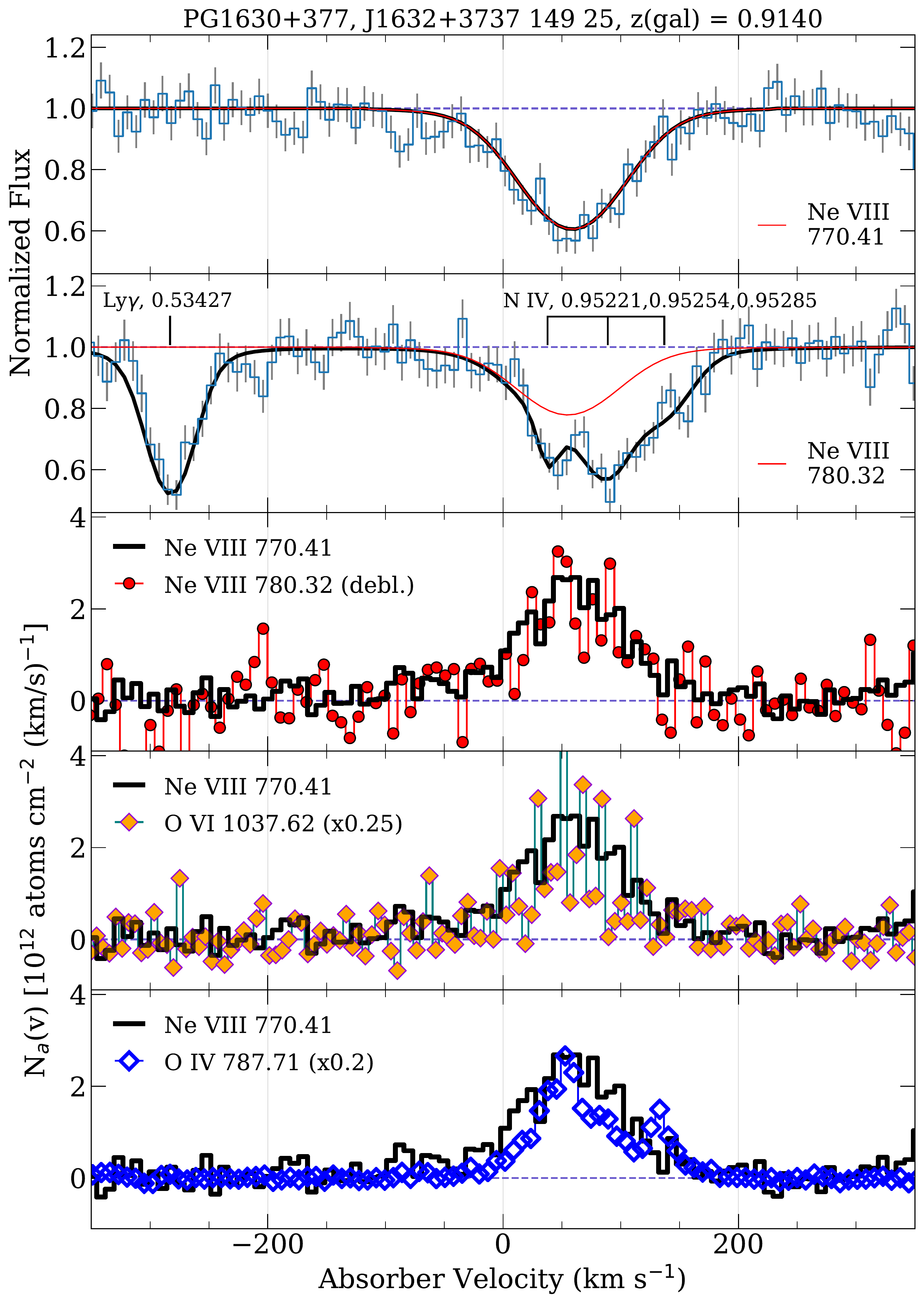}
\caption{continued}
\end{figure*}

\section{Appendix: \neeight\ line identifications}
Here, we present additional information on our identification of \neeight\ absorption within the CASBaH data.  The wavelengths of the Ne~\textsc{viii} doublet and the minimum redshift required to redshift the lines into the {\it HST} bandpass inevitably leads to some blending with lines from other systems, but this blending is surmountable.  In addition, the Ne~\textsc{viii} lines are relatively weak.  For comparison, we note that in the Sun, the abundance of neon is $\approx 6\times$ lower than the abundance of oxygen \citep{Asplund:2009rf}, and the atomic transition strengths of the Ne~\textsc{viii} lines are $\approx 2\times$ weaker than those of \textsc{O~vi} as well \citep{Verner:1994lr}.  Consequently, identification of Ne~\textsc{viii} requires high S/N and careful examination of the data.


As shown in Figure \ref{fig:stackplots}, all of the CGM \neeight\ absorbers that we have identified are detected in other lines such as \hone\ \textsc{C~iii}, \textsc{O~iv}, and \osix, often with multiple components. Though we have searched for Ne~\textsc{viii} doublets that do not have other metals and/or \textsc{H~i} at the same redshift, we have found no compelling examples of ``Ne~\textsc{viii}-only'' intervening absorbers; the interveing Ne~\textsc{viii} always exhibit other metals such as \textsc{O~iv} and \textsc{O~vi}. Correspondence between the velocity component structures of \neeight\ and of these other lines can lend further credence to the \neeight\ identifications, because the chance alignment of unrelated interloping profiles is  highly unlikely (we conservatively quantify this likelihood below).  

Figure \ref{fig:ne8evidence} presents the detailed absorption data that we have used to identify the CGM \neeight\ absorbers in our sample.  Here, the histograms in the top two panels show the continuum-normalized 
\neeight\ $\lambda\lambda$ 770.41, 780.32 \AA\ absorption profiles (also shown in Figure \ref{fig:stackplots}) with 1$\sigma$ flux uncertainties. On these normalized profiles, we overplot our overall Voigt-profile fits (including blends with interloping lines from other redshifts or the Milky Way) with smooth black lines, and we show the components for the Ne~\textsc{viii} only with smooth red lines; annotations in these panels indicate the identification of interloping lines. The third panel shows apparent column density profiles \citep{Savage:1991vn} of each \neeight\ line of the doublet, and the remaining panels show apparent \neeight\ column density profiles overplotted with those of lines from other species identified at the same redshift.   In cases where the line of interest (from \neeight\ or otherwise) was in a blended complex, we have deblended the interloping lines based on the Voigt profile models (which include all relevant lines elsewhere in the spectra).  From these data, we see that the spectra show consistent alignment and optical depth from the two lines of the Ne~\textsc{viii} doublet, and we see that the Ne~\textsc{viii} apparent column-density profiles are often very similar to the profiles of other metals in the absorbers.

Having shown that our identified \neeight\ components not only align in centroid with lines from other species in the same absorption system but also have close correspondence over the \textit{shape of the profile}, we now quantify the probability of interloping absorption occurring at redshifts such that they could mimic associated \neeight\ components.  We reemphasize that we produced identifications for \textit{all} lines near the would-be \neeight\ locations for all systems in our sample; these lines include \hone\ Lyman series lines and metal lines from other redshifts, and they have been included in the Voigt profile fits.  Thus, in most cases the only possible interloping line is \hone\ \lya\ at $z<0.323$ (the redshift of a putative Ly$\alpha$ line at the observed wavelength of our highest redshift \neeight\ system).  To evaluate the probability of two interloping Ly$\alpha$ lines mimicking the Ne~\textsc{viii} doublet, we conducted a Monte Carlo simulation experiment to estimate the likelihood of such chance alignments.  We consider such an occurrence as two \lya\ lines falling at redshifts where a) the lines would have the correct spacing in observed wavelength to mimic the \neeight\ $\lambda\lambda$ 770, 780 \AA\ doublet and b) the redshift of this imposter `\neeight' doublet matches the redshift of another absorption system in the sightline.  

Our procedure for this experiment was as follows:  We first generated a mock sample of $z<0.323$ \lya\ forest absorbers using random draws from the \hone\ column density distribution function of \cite{Danforth:2016aa}; our range of column densities was set by the column densities corresponding to the equivalent widths implied by our weakest and strongest detected \neeight\ absorption components (log N(\hone)/cm$^{-2}$ = 12.3 - 13.3).  This yielded 70-90 mock \lya\ components per `sightline'.  Then, we calculated the redshifts at which each line of the \neeight\ doublet would fall at the observed wavelengths of the mock \lya\ lines.   Assuming each mock \lya\ was a 
\neeight\ $\lambda$ 770 \AA\ line, we tested whether any other mock \lya\ lines fell within 30 km s$^{-1}$ \citep[commensurate with reported COS wavelength solution uncertainties;][]{Burchett:2015rf,Wakker:2015rf} of the would-be location of \neeight\ $\lambda$ 780 \AA; such matches were flagged as potential \neeight\ imposters.  Because all of our \neeight\ absorbers also exhibit absorption from other metal ions, we also produced randomized instances of mock metal-line systems by drawing ten random redshifts from the observed redshift range of our \neeight\ absorbers.  This number of metal-line absorbers is based on the average number of actual metal-line systems identified within the CASBaH sightlines at the relevant redshifts.  Finally, we considered any of the potential fake doublets aligned within 50 km s$^{-1}$ of a mock metal system to be a legitimate imposter.  Repeating this procedure for 2000 mock sightline realizations, we found a contamination rate of 0.5\% for \lya\ lines posing as \neeight\ doublets associated with other metal line systems.

This \lya\ contamination estimate is conservative for several reasons.  First, we place no constraints on the relative line strengths of the fake \neeight\ doublets; the $\lambda$ 770 \AA\ line should be $\sim 2\times$ stronger than $\lambda$ 780 \AA, but we simply test the redshift spacing of the masquerading \lya\ lines.  Second, for several of our systems, we detect multiple \neeight\ components that align with the component structure seen in other metal species; the Monte Carlo simulations were not required to match this detailed component structure.  Inclusion of both of these factors would substantially reduce the probability of randomly distributed Ly$\alpha$ lines mimicking the Ne~\textsc{viii} doublets that we have identified.

\bibliographystyle{apjMOD}

\end{document}